\documentclass[structabstract]{aa1}  
\usepackage{graphicx}
\usepackage{txfonts}
\usepackage{natbib}
\bibpunct{(}{)}{;}{a}{}{,} 
\usepackage{color}

\begin{document}
   \title{Deep wide-field near-infrared survey of the Carina Nebula \thanks{Based on observations collected with the HAWK-I instrument at the VLT at Paranal Observatory, Chile, under ESO program 60.A-9284(K).}}

   \author{T.~Preibisch\inst{1} 
          \and
          T.~Ratzka\inst{1}
          \and
          B.~Kuderna\inst{1}
          \and
          H.~Ohlendorf\inst{1}
          \and
          R.~R.~King\inst{2}
          \and
          S.~Hodgkin\inst{3}
          \and
          M.~Irwin\inst{3}
          \and
          J.R.~Lewis\inst{3}
          \and
          M.J.~McCaughrean\inst{2,4} 
          \and
          H.~Zinnecker\inst{5,6,7}
          }

   \institute{
              Universit\"ats-Sternwarte M\"unchen,
              Ludwig-Maximilians-Universit\"at,
              Scheinerstr.~1, D-81679 M\"unchen, Germany\,\,\,
              \hfill \email{preibisch@usm.uni-muenchen.de}
         \and
             Astrophysics Group, College of Engineering, 
            Mathematics and Physical Sciences, University of Exeter, 
            Exeter EX4 4QL, UK
          \and
             Cambridge Astronomical Survey Unit,
             Institute of Astronomy, Madingley Road, Cambridge, CB3 0HA, UK
         \and
             European Space Agency, Research \& Scientific Support Department, 
             ESTEC, Postbus 299, 2200 AG Noordwijk, The Netherlands
         \and
             Astrophysikalisches Institut Potsdam, An der 
             Sternwarte 16, D-14482 Potsdam, Germany
             \and
   Deutsches SOFIA Institut, Universit\"at Stuttgart, Pfaffenwaldring 31,
    70569 Stuttgart, Germany
    \and
NASA-Ames Research Center,
MS 211-3, Moffett Field, CA 94035, USA
             }

\titlerunning{Deep wide-field near-infrared survey of the Carina Nebula}
\authorrunning{Preibisch et al.}

   \date{Received 25 February 2011 / Accepted 4 April 2011}

 
  \abstract
   {The
Great Nebula in Carina is a giant \ion{H}{II} region and a superb location 
in which to study the physics of violent massive star formation, but
the population of the young low-mass stars
remained very poorly studied  until recently.
   }
   {
Our aim was to produce 
a near-infrared survey 
that is deep enough
to detect the full low-mass stellar population 
(i.e.~down to $\approx 0.1\,M_{\odot}$ and for extinctions up to
$A_V \approx 15$~mag) and wide enough 
to cover all important parts of the Carina Nebula complex \textbf{(CNC)}, 
including the clusters
Tr~14, 15, and 16 as well as the South Pillars region.
   }
   {
We used HAWK-I at the ESO VLT to survey 
of the central $\approx 0.36\,\rm deg^2$~ area 
of the Carina Nebula. These data reveal 
more than 600\,000 individual infrared sources down to
magnitudes as faint as
$J \approx 23$, $H \approx 22$, and $K_s \approx 21$.
The results of a recent deep X-ray survey (which is complete down to
stellar masses of $\sim 0.5 - 1\,M_\odot$) are used to distinguish
between young stars in Carina and background contaminants.
We analyze color-magnitude diagrams (CMDs)
to derive information about the ages and masses of the low-mass stars.
   }
   {
The ages of the low-mass stars agree 
with previous age estimates for the massive stars.
The CMD suggests that $\approx 3200$ of the X-ray selected stars have
masses of
$M_\ast \ge 1\,M_\odot$; this number is in good agreement with extrapolations
of the field IMF
based on the number of high-mass ($M_\ast \ge 20\,M_\odot$) stars
and shows that there is no deficit of low-mass stars in the CNC.
The HAWK-I images confirm
that about 50\% of all young stars in Carina are in a widely distributed,
non-clustered spatial configuration.
Narrow-band images reveal six molecular hydrogen emission objects 
(MHOs) that
trace jets from embedded protostars.
However, none of the optical HH objects shows molecular hydrogen
emission, suggesting that the jet-driving protostars are located
very close to the edges of the globules in which they are embedded.
   }
   {
The near-infrared excess fractions for the stellar population in Carina are
lower than typical for other,
less massive clusters of similar age, suggesting that
the process of circumstellar disk dispersal proceeds on
a faster timescale in the CNC than in the more quiescent regions,
most likely due to the very high level of massive star
feedback in the CNC.
The location of all but one of the known jet-driving protostars at the edges
of the globules adds strong support to the scenario that their formation
was triggered by the advancing ionization fronts.
}


   \maketitle
%

\section{Introduction}

Most stars in the Galaxy are thought to be born in massive star-forming regions 
\citep[e.g.,][]{Blaauw64,MS78,Briceno07},
and therefore
{\em in close proximity to massive stars}.
This also applies to the origin of our
solar system, for which recent investigations found convincing
evidence that it formed in a large cluster, consisting of
(at least) several thousand stars, and that the original solar
nebula was directly affected by nearby massive stars
\citep[e.g.,][]{Adams10}.
As a consequence, the
role of environment is now an essential topic in studies
of star  and planet formation.
The presence of high-mass stars can lead to  physical conditions
that are vastly different from those in regions where
only low-mass stars form.
The very luminous O-type stars 
profoundly influence their environment by their strong
ionizing radiation, powerful stellar winds, and, finally,
by supernova explosions.
This feedback can disperse the sur\-round\-ing natal molecular clouds
\citep[e.g.,][]{Freyer03}, 
and thus terminate the star formation process (= negative feedback). 
However, ionization fronts 
and expanding superbubbles can also compress near\-by clouds and 
thereby trigger the formation of new generations of stars
\citep[= positive feedback; e.g.,][]{Gritschneder10,PZ07}. 
These processes determine the key outputs from star
formation, including the stellar mass function, the total star formation
efficiency, as well as the evolution of circumstellar disks
around the young stars \citep[e.g.,][]{Clarke07} and the frequency of planetary formation.
While this general picture is now well established,
the details of the feedback processes, i.e.~cloud dispersal
on the one hand, and triggering of star formation on the other
hand, are still only poorly understood. 
A fundamental problem is that
nearly all massive star-forming clusters 
with high levels of feedback are quite far away and therefore
difficult to study.
At distances greater than 5~kpc,  the detection and characterization
of the {\em full} stellar populations 
is very difficult (if not impossible); often, only the bright 
high- and intermediate-mass stars can be studied,
leaving the low-mass ($M \le 1\,M_{\odot}$) stars 
unexplored or even undetected.
However, since the low-mass stars constitute the vast majority
of the stellar population, a
good knowledge of the low-mass stellar content is essential
for any inferences on the initial mass function and
for understanding the nature of the star formation process.
Detailed information about the low-mass stars and their
protoplanetary disks is also crucial for investigations
of the interaction between the high- and
low-mass stars.
The intense UV radiation from massive stars
may remove considerable amounts of the circumstellar material
 from nearby young stellar objects (YSOs, hereafter). This may limit
the final masses of the low-mass stars
\citep[see][]{Whitworth04} and should also affect
the formation of planets \citep[see, e.g.,][]{Throop05}.

In this context, the
Great Nebula in Carina \citep[NGC 3372; see, e.g.,][for an overview]{SB08}
provides a unique target for studies of massive star feedback.
At a very well known and moderate distance of 2.3\,kpc, 
the Carina Nebula Complex (CNC, hereafter) represents the nearest 
southern region with a large massive stellar population \citep[65 
O-type stars; see][]{Smith06}. 
Among these are several
of the most massive  ($M \ga 100\,M_\odot$) and luminous stars 
known in our Galaxy, e.g., the famous Luminous Blue Variable $\eta$\,Car, 
the O2~If* star HD~93129Aa, several O3 main sequence stars and Wolf-Rayet  stars.
The CNC is \textit{the most nearby region that samples the top of the 
stellar mass function}.
The presence of stars with $M \ga 100\,M_\odot$ implies
that the level of feedback in the CNC is already close to that in 
more extreme extragalactic starburst regions, while at the same
time its comparatively moderate distance guarantees that we still
can study details of the cluster and cloud structure at
good enough spatial resolution and detect and characterize
the low-mass stellar populations.
Due to this unique combination of properties, the CNC represents
the best galactic analog of giant extragalactic
\ion{H}{II} and starburst regions.

Most of the very massive stars in the CNC 
reside in several loose clusters, including Tr~14, 15, and 16,
for which have ages between $\sim 2 $ and $\sim 8$~Myr have been
estimated \citep[see][]{SB08}.  
The CNC is thus often denoted as a ``cluster of clusters''.
In the central region,
the molecular clouds  have already been largely dispersed by the 
feedback from the numerous massive stars. Southeast of $\eta$~Car, in 
the so-called ``South Pillars'' region,
the clouds are
eroded and shaped by the radiation and winds from
$\eta$~Car and Tr~16, giving
rise to numerous giant dust pillars, which feature very prominently 
in the mid-infrared images made with the {\it Spitzer} Space Observatory
\citep{Smith10b}.
The complex contains more than $10^5\,M_\odot$ of gas and dust
 \citep[see][]{Yonekura05,Preibisch_Laboca},
and deep infrared observations 
show clear evidence of ongoing star formation in these clouds.
Several deeply embedded YSOs 
\citep[e.g.,][]{Mottram07} and a spectacular young cluster
\citep[the ``Treasure Chest Cluster''; see][]{Smith05} have been 
found. 
 A deep {\it HST} H$\alpha$ imaging survey revealed
dozens of jet-driving YSOs \citep{Smith10a}, and
{\it Spitzer} surveys located numerous embedded intermediate-mass
protostars  throughout the
Carina complex \citep{Smith10b,Povich11a}. 
The formation of this substantial population of very young
($\leq 1$~Myr), 
partly embedded stars was probably triggered by the
advancing ionization fronts that originate from the
(several Myr old) high-mass stars.
The CNC provides an excellent target to
investigate these effects
on the formation and evolution of low-mass stars (and their forming
planetary systems) during the first few Myr.


While the un-obscured population of high-mass stars $(M \geq 20\,M_\odot)$
in the CNC
is well known and characterized, 
the (much fainter) low-mass $(M \leq 1\,M_\odot)$ stellar 
population remained largely unexplored until now. 
One reason for this is the faintness of the 
low-mass stars: at a distance of 2.3~kpc, an
age of 3~Myr, and assuming typical extinctions of $A_V = 3.5$~mag \citep{Preibisch_CCCP},
stars with masses of $0.5\,M_\odot\;[0.1\,M_\odot]$ are predicted to
have magnitudes of $V=23.6\,[26.6]$, $J=17.0\,[19.2]$, and
$K=15.5\,[17.7]$ \citep[see][]{Baraffe98}. 
The existing near-infrared (NIR) observations were either too shallow to 
detect the full low-mass population \citep[e.g.,][]{Sanchawala07},
or covered only very small parts of the complex \citep[e.g.][]{Ascenso07}.

This was the motivation for the deep wide-field NIR
survey presented in this paper. 
Our survey is $\ge 5$~mag deeper than the 2MASS data, deep
enough to detect essentially all young stars
in the CNC, and 
can be used to derive information about the ages and masses of the
young low-mass stars. 
A fundamental restriction is that 
the NIR data alone do not allow us to distinguish 
young stars in the  CNC from
the numerous galactic field stars in the area (see discussion in Sect.~2).
For this purpose, we use the recent results from the
{\it Chandra} Carina Complex Project (CCCP), that has mapped
the entire extent of the CNC ($\approx 1.4\,\rm deg^2$) in X-rays. 
A complete overview of the CCCP can be found in
\citet{Townsley11} and associated papers in a 
Special Issue of the {\em Astrophysical Journal Supplements}.
The CCCP data led to the detection of 14\,368 individual X-ray sources.
Based on a statistical analysis of the properties of
each individual source, \citet{Broos11b} classified 
10\,714 of these as young stars in the CNC.
The identification of HAWK-I infrared counterparts of the
X-ray sources in the CNC and a few basic aspects of the X-ray
selected population of young stars have been recently presented
in \citet{Preibisch_CCCP}.
In this paper we discuss the complete set of HAWK-I data (Sec.~2) and
present a detailed analysis of the NIR color-magnitude
diagrams (Sec.~3). In Section 4 we analyze the NIR
excess fractions for the different clusters in the CNC.
In Section 5 we
draw global implications 
on the size of the stellar population and the IMF of the CNC.
Section 6 contains the results of our search for protostellar
jets in  H$_2$ narrow-band images. In Sec.~7 we look at the
spatial distribution of the young stars and list
newly discovered stellar clusters in the CNC.
Global conclusions on the star formation process in the CNC
are discussed in Sec.~\ref{conclusions.sec}.


\section{HAWK-I observations and data analysis}

\begin{figure*}
\centering
\includegraphics[width=14.0cm]{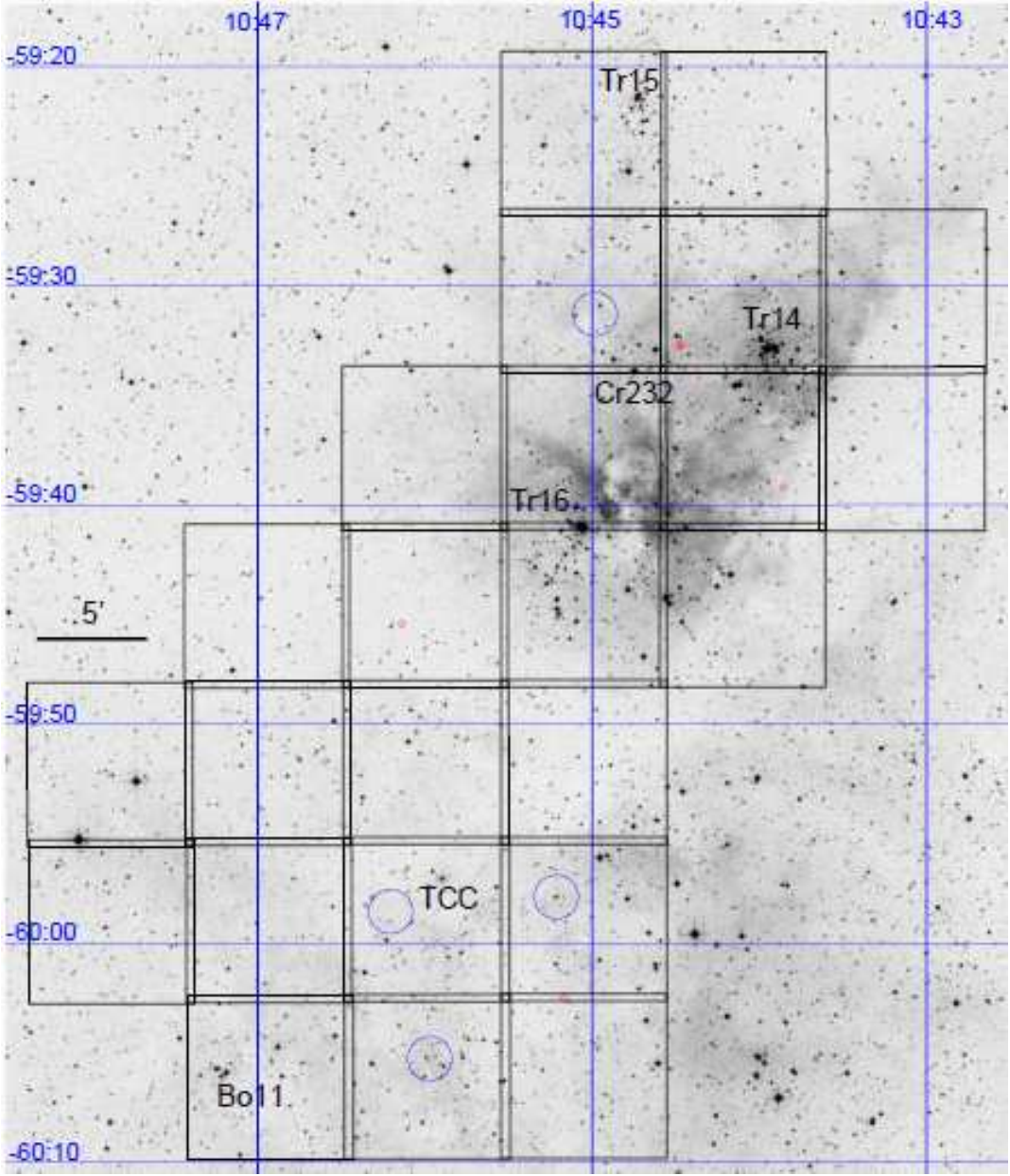}
\caption{Negative gray scale representation of an optical image of the Carina Nebula
 taken from the red Digitized Sky Survey plates, with superposed outline
of the HAWK-I mosaic pattern.
The individual HAWK-I mosaic fields are shown as boxes with a size
of $7.5' \times 7.5'$. The clusters Tr~14, Tr~15, Tr~16, Cr~232, Bo~11,
and the Treasure Chest Cluster (TCC) are marked.
The $2'$ diameter blue circles mark the positions of the four embedded clusterings
discussed in Sect.~\ref{clusters.sect}, and the six small red circles 
indicate the positions of the six Molecular Hydrogen Emission Objects
discussed in \ref{mhos.sec}.
North is up and east to the left. A grid of J2000 coordinates is also shown.
              \label{hawki.fig}%
    }
\end{figure*}

\begin{figure*}
\centering
\includegraphics[width=14.0cm]{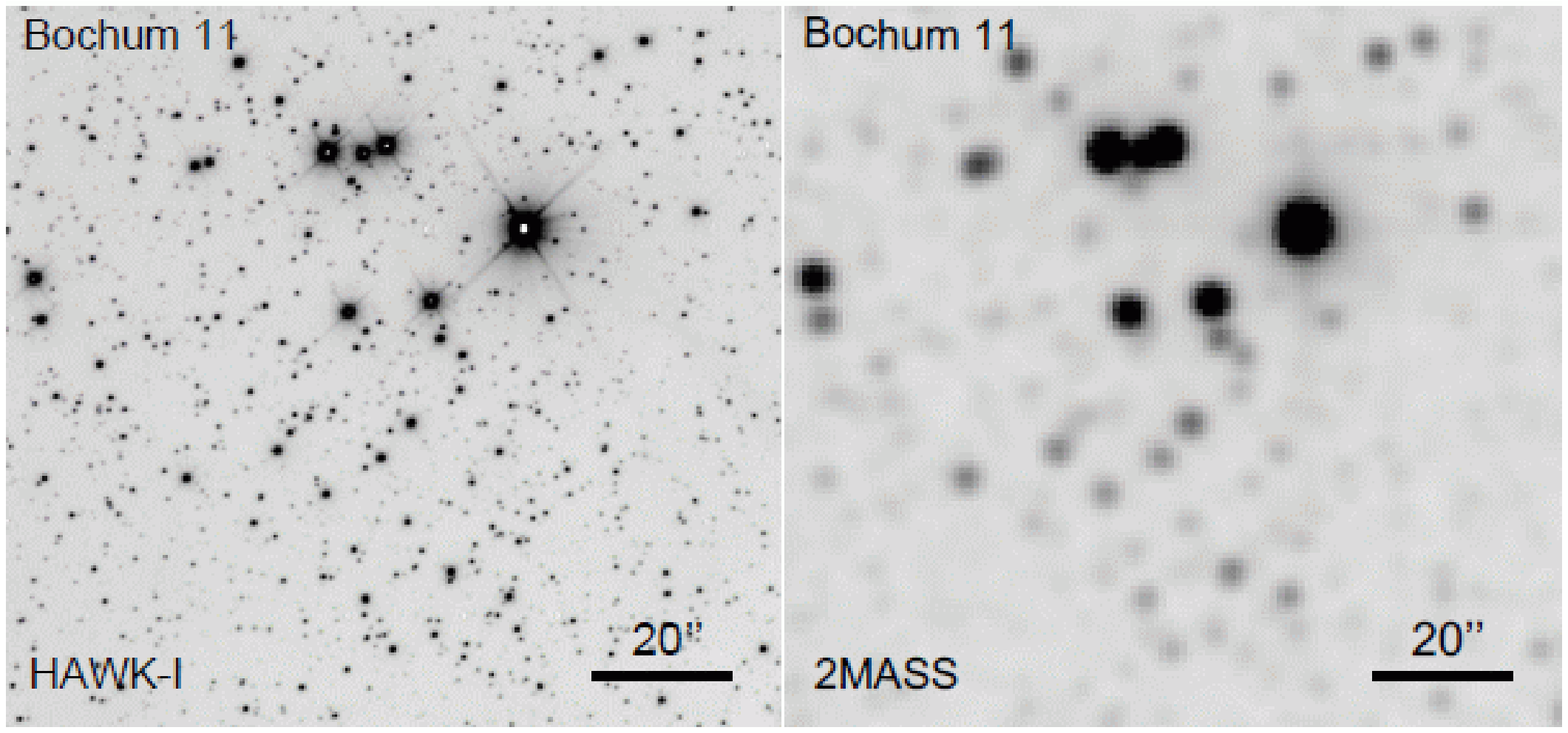}
\includegraphics[width=14.0cm]{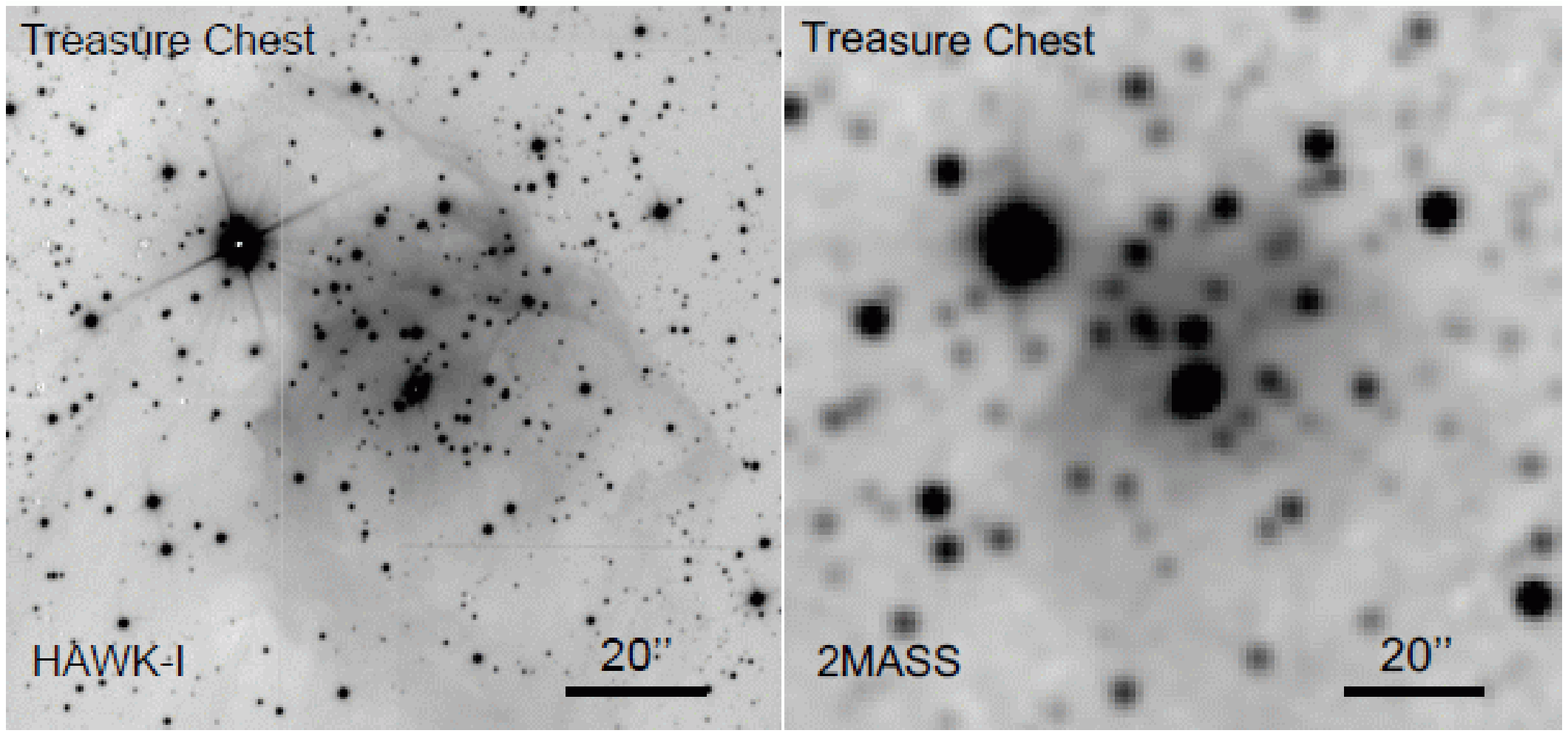}
\caption{Comparison of HAWK-I (left) and 2MASS (right) $H$-band
images of Bochum 11 (above) and the Treasure Chest cluster (below).
              \label{bo11.fig}%
    }
\end{figure*}

The NIR imager HAWK-I 
\citep[see][]{HAWKI08} at the 
ESO 8\,m Very Large Telescope 
is equipped with a
mosaic of four Hawaii 2RG $2048\times 2048$ pixel detectors with
a scale of $0.106''$ per pixel.
The camera
has a field of view on the sky of $7.5' \times 7.5'$
with a small cross-shaped gap of $\sim 15''$ between the four
detectors. 
The observations of the CNC were performed 
from 24~to 31~January 2008 in service mode
as part of the scientific
verification program for HAWK-I.
In addition to the standard
broadband $J$, $H$, and $K_{\rm s}$ filters, 
we used  the H$_2$ 
($2.109-2.139\,\mu$m) and Br\,$\gamma$ ($2.150-2.181\,\mu$m)
narrow-band filters.

Our survey consists of a
mosaic of 24 contiguous HAWK-I fields,
covering a total area of about 1280 square-arcminutes; it 
includes the central part of the Nebula with $\eta$~Car and Tr~16, 
the clusters Tr~14 and Tr~15, and large 
parts of the South Pillars (see Fig.~\ref{hawki.fig}).

The total observing times  for each mosaic position  were
12, 8, and 5 minutes in $J$, $H$, and $K_{\rm s}$.
Additional 5 minutes were spent with each of the 
H$_2$ and Br\,$\gamma$ narrow band filters.
In order to account for the gaps of the detector array,
we used a 5-point dither pattern with offsets of 
(0,0), ($-40''$, $+40''$), ($-40''$, $-40''$), ($+40''$, $-40''$),
 and ($+40''$, $+40''$)
at each individual mosaic position.
Details of the observing parameters  are listed
in Table~\ref{obs.tab}.

\begin{table}
\caption{Observing parameters
}
\label{obs.tab}       
\begin{tabular}{l|rccc}
Filter &  Offsets   &   Integrations & Integration & Total exposure\\
      &       &                &  times [s]     & time [s]\\\hline
 $J$ &  $2\times5$  &  24 &  3 & 720\\
 $H$ &   $2\times5$  &  16 &  3 & 480\\
 $K_{\rm s}$ &  5   &  20 &  3 & 300\\
 $H_{\rm 2}$ &  5   &  2 &  30 & 300\\
 $Br\,{\gamma}$ &  5   &  2 &  30 & 300\\
\end{tabular}
\end{table}

Thanks to the very good seeing conditions during the observations,
most of the HAWK-I images are of a very high image quality.
The FWHM of the Point-Spread-Function in the HAWK-I images is typically 
$0.6'' - 0.8''$. 
This good image quality led to the detection of
interesting structures such as the
edge-on circumstellar disk described in 
\citep{Preibisch11c}.

As examples for the quality of the HAWK-I data, we show 
comparisons of the
HAWK-I and 2MASS images of selected regions in the CNC in Fig.~\ref{bo11.fig}.

\subsection{Photometry and construction of the HAWK-I source catalog}

The procedures performed to derive photometry from the HAWK-I images
and the
construction of the source catalog
is described in detail in \cite{Preibisch_CCCP}. Here we
briefly summarize the main points.
All HAWK-I data were processed and calibrated by the Cambridge
Astronomical Survey Unit using pipeline software as described in
\citet{Irwin04}.
A photometric analysis was performed for all images obtained
in the three broad-band $J$, $H$, and  $K_s$, 
following closely the
procedures described in detail in \citet{Hodgkin09}. 
The photometric calibrators are drawn from stars in the 
2MASS Point Source Catalog, which are
present in large
numbers (several hundred) in each HAWK-I mosaic field.
The source lists of the individual mosaic positions were then combined
into total catalogs (per band) by  removing duplicates from the overlap
regions; this was achieved by keeping the source with the
 highest signal-to-noise ratio. The final 
catalog was generated by merging the source 
catalogs for the individual bands into a combined catalog.
The maximum allowed spatial offset for inter-band identifications
of individual objects was set to $\le 0.3''$.

The final HAWK-I photometric catalog contains 600\,336 individual objects.
Most (502\,714) catalog objects are simultaneously detected
 in the $J$-, $H$-, and the $K_s$-band.
Objects as faint as
$J \sim 23$, $H \sim 22$, and $K_s \sim 21$ are 
detected with S/N~$\ge 3$.
Typical values for the completeness limit across the field  are
$J_{\rm compl} \sim 21$, $H_{\rm compl} \sim 20$, 
and $K_{s,\, \rm compl} \sim 19$; nearly all objects above these limits 
are S/N~$\ge 10$ detections.
Our survey is thus about 5~mag deeper than 2MASS and represents
{\em the largest and deepest NIR survey of the
CNC obtained so far}.

The absolute photometric accuracy of the catalog was characterized
by a comparison of  the HAWK-I and 2MASS photometry for a 
sample of more than 700 stars. The standard deviations between 
HAWK-I and 2MASS magnitudes and colors are found to be
$\sigma_J = 0.11$~mag, $\sigma_H = 0.08$~mag, $\sigma_{K_s} = 0.11$~mag,
$\sigma_{J-H} = 0.12$~mag, and $\sigma_{H-K_s} = 0.11$~mag.
The accuracy is mainly limited by
the photometric quality of the 2MASS stars suitable
for comparison and the
strong and highly variable diffuse nebulosity in the area of the
HAWK-I mosaic.

\subsection{Background contamination in the HAWK-I catalog\label{imf_expectation}}

Due to the CNC's location almost exactly on the galactic plane,
the degree of field star contamination is very high. This
causes a fundamental limitation of the usefulness of the infrared source catalog.
The degree of contamination was recently quantified in two
{\em Spitzer} studies of the CNC. 
\cite{Smith10b} and \citet{Povich11a} found that only 
$\approx 3\%$ of the $\sim 50\,000$ {\em Spitzer} infrared sources in the
CCCP survey area can be classified as candidate YSOs.
The contamination rate of our deep NIR HAWK-I data is presumably 
similarly high.
This suggests that {\em the vast majority  ($\gtrsim 90\%$)
of the
$\sim 600\,000$ detected infrared sources in our HAWK-I catalog
must be foreground or background sources, unrelated to the CNC}.

\subsection{Contamination and incompleteness of NIR excess selected samples}

\begin{figure} \centering
\includegraphics[width=8.0cm]{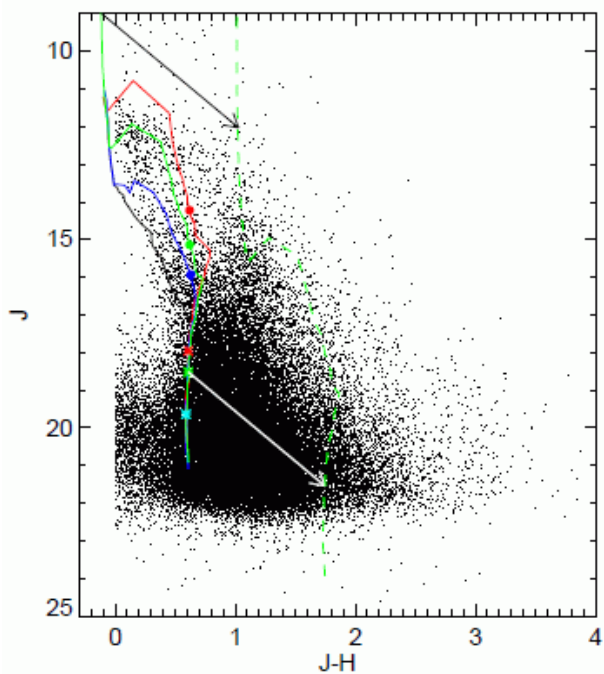}
\caption{Color-Magnitude Diagram of all HAWK-I sources with NIR
excesses. The solid lines show isochrones for ages of
1~Myr (red), 3~Myr (green), 10~Myr (blue), and the ZAMS
composed from the models of \citet{Baraffe98} for the mass range
0.02 to 0.5~M$_\odot$ and \citet{Siess00} for the
mass range 0.5 to 7~M$_\odot$.
The large solid dots mark the positions of $1\,M_\odot$ stars,
the asterisks those of $0.075\,M_\odot$ objects on the 1, 3, and 10~Myr
isochrones.  
extinction vector for $A_V = 10$~mag. 
The arrows indicate reddening vectors for $A_V = 10$~mag and were
computed for an extinction law with $R_V = 4$  \citep[see][]{Povich11b}.
The dashed green line
shows the 3~Myr isochrone reddened by that amount.
Note that all bright objects with  $J < 12.5$   are in the saturation/non-linearity
regime of the HAWK-I data and their photometry is therefore unreliable.
\label{cmd-ex.fig}}
\end{figure}

This high degree of contamination requires a very efficient and reliable
selection method to identify YSOs among the very numerous contaminants.
A widely used method is to select 
objects with infrared excess emission, 
which is a tracer of circumstellar material, as YSO candidates.      
In order to test whether such an approach could be useful for our data,
we analyzed the
$J\!-\!H$ vs.~$H\!-\!K_s$ color-color diagram of the HAWK-I sources.
Objects are classified as NIR excess sources if
they lie at least $\geq 1\sigma_{\rm phot}$ and more than 0.05~mag to the 
right and below the reddening band, 
based on the intrinsic colors of dwarfs \citep{BB88}
and an extinction vector with slope 1.73, and above $J\!-\!H=0$. In this way,
50\,950 of the 502\,714 HAWK-I sources with complete $J$, $H$, and $K_s$ photometry
are classified as NIR excess sources.
The color-magnitude diagram (CMD) of these NIR excess sources  is shown in
Fig.~\ref{cmd-ex.fig}.

However, there are two reasons why we think that this excess-selected
sample is \textit{not} a useful sample of the YSOs in the CNC.
First, due to the very high level of background contamination,
even an excess selected sample will still be highly polluted by
background objects. Several recent spectroscopic
studies of infrared sources in other star forming regions have shown
that  excess-selected YSO candidate samples 
can be {\em strongly contaminated}
by background sources such as
evolved Be stars, carbon stars, or planetary nebulae, which often show
NIR colors very similar to those of YSOs
\citep[e.g.,][]{Mentuch09,Oliveira09,Rebull09}.
Our CMD suggests this to be the case here too:
a large majority (78\%) of the NIR excess objects is 
found at very faint magnitudes ($J > 19$). 
Comparing the location of the sources in the CMD
to pre-main sequence models shows that
61\% of all NIR excess objects are below the expected reddening 
vector for 3~Myr old $0.075\,M_\odot$ objects in Carina.
While some of these faint objects may be deeply embedded YSOs,
stars with edge-on circumstellar disks, or disk-bearing young brown dwarfs,
the majority are very likely \textit{not} young stars in the CNC but
background objects.
Furthermore, we note that the
HAWK-I images are deep enough to detect numerous extragalactic
objects. Since star-forming galaxies and AGN often
display NIR colors similar to those of YSOs,
this leads to even higher contamination
rates in NIR excess selected samples of YSO candidates\footnote{
Based on the number counts of star-forming galaxies in the data from the
UKIRT Infrared Deep Sky Survey data by \cite{Lane07} we can expect 
the presence of about 4700 star-forming galaxies with $K_s \leq 20$
in the HAWK-I field. From a purely statistical point-of-view,
all of the 2157 NIR excess objects
with $K_s > 20$ in the HAWK-I catalog may well be extragalactic sources.}.

The second problem of a NIR excess selected YSO candidate sample
is its \textit{incompleteness}.
It is well known that
NIR excess emission in young stars 
disappears on timescales of just a few Myr \citep[e.g.,][]{Briceno07};
at an age of $\sim 3$~Myr, 
only $\sim 50\%$ of the young stars still
show NIR excesses, and by $\sim 5$~Myr this is reduced to $\sim 15\%$.
Since the expected ages of most young stars in the CNC are
several Myr, {\em any excess-selected YSO sample will be 
highly incomplete}.

Considering these problems, we conclude that
due to the very high degree of background contamination 
and the high expected level of incompleteness of
NIR excess selected YSO candidate samples,
the NIR data alone cannot provide a reasonably clean and complete sample of 
YSOs.
Without additional information, the HAWK-I data alone can yield only very 
limited insight into the young stellar populations in the CNC.

\subsection{Combination of infrared and X-ray data}

Sensitive X-ray observations can provide
a very good solution for the problem of identifying the
young stars in an extended complex such as the CNC.
X-ray surveys
detect the young stars by their strong X-ray emission
\citep[e.g.,][]{Feigelson07}
and efficiently
discriminate them from the numerous
older field stars in the survey area.
X-rays are equally sensitive to young stars which have already
dispersed their circumstellar disks, thus avoiding the bias
introduced when selecting
samples based only on infrared excess.
Many X-ray studies of star forming regions have demonstrated
the success of this method
\citep[see, e.g.,][]{PZH96,PZ02,Broos07,coronet,Wang10}.
Also, the
 relations between the X-ray properties and basic stellar
properties in young stellar populations
are now  very well established from very deep X-ray
observations such as the 
{\it Chandra} Orion Ultradeep Project 
\citep[see][]{Getman05,Preibisch_coup_orig}.

The
combination of the \mbox{HAWK-I} infrared source catalog
with the list of X-ray sources detected was an essential
aspect of the \textit{Chandra} Carina Complex Project \citep[CCCP; see][]{Townsley11}. 
The details of the infrared -- X-ray
source matching is described in \citet{Preibisch_CCCP}, and
here we just briefly summarize the main aspects:
The area of the \mbox{HAWK-I} mosaic
covers 27\% of the CCCP survey area
and contains 52\% of all 14\,368 detected X-ray sources. 
The \mbox{HAWK-I} infrared catalog provided infrared matches
for 6583 of these X-ray sources.
After adding matches from the 2MASS catalog for very bright stars
that are saturated in the \mbox{HAWK-I} images, infrared counterparts
were established for 6636 X-ray sources.
6241 of these have valid
photometry in all three of the $J$-, $H$-, and $K_s$-bands,
and 6173 (93.0\%) are classified as Carina members \citep[see][]{Broos11b}.

Using the NIR photometry of the X-ray selected Carina members, it was found
that the typical extinction values for the diskless stars 
range from $A_V \sim 1.6$~mag to $\sim 6.2$~mag (central 80\% percentile). 
While this clearly shows a
considerable range of differential extinction between individual stars
in the complex, it also implies that the 
vast majority of the X-ray detected young stellar
population in the CNC shows rather moderate extinction, 
$A_V \le 10$~mag.
The X-ray selected Carina members can thus be considered as a ``lightly obscured''
population of young stars.
As will be discussed in Sec.~\ref{conclusions.sec}, deep \textit{Spitzer}
data suggest the presence of an additional (but smaller)
 population of embedded YSOs
which are {\em not} detected in the CCCP X-ray data.

%
\section{Color-Magnitude Diagrams for selected clusters in the CNC}
%

The stars in the CNC are distributed in a number
of clusters (most notably the prominent optical clusters Tr~14, 15, and 16)
 and a widely dispersed population of young stars.
In this section we consider the properties of the stellar populations
in these different parts of the CNC. 
In order to avoid the background contamination problem, we
restrict our analysis to the X-ray selected members, using 
the results of the clustering analysis of
\citet{Feigelson11}.

In our analysis of the CMDs 
we use the isochrones derived from the pre-main
sequence models of \citet{Baraffe98} for the mass range
0.02 to 0.5~M$_\odot$ and those of \citet{Siess00} for the
mass range 0.5 to 7~M$_\odot$.
For the more massive stars, which are on the main sequence
or already in their post-main sequence phases,
we used the Geneva stellar models from \citet{Lejeune01} (model
 iso-c020-0650) for the mass range 7 to 70 ~M$_\odot$.

Before we consider the individual CMDs, we want to point out that there
are fundamental limitations to 
the accuracy of stellar mass- and age-determinations from CMDs 
(or HRDs); 
 different studies of the same stellar population can sometimes produce
severely discrepant results \citep[see discussion in][]{Hartmann01,Hartmann03},
especially with respect to the presence or absence of age spreads.
Factors such as differential extinction, photometric variability, overestimates
of the stellar luminosity due to unresolved binary companions,
and the effects of accretion\footnote{\cite{Baraffe09}
showed that stars with intermittently variable
accretion rates can produce a large spread of luminosities
such that a population of stars with identical ages of a few Myr
may be wrongly interpreted to have an age spread of as much
as $\sim 10$~Myr.}
can cause substantial scatter in the diagram 
\citep[see, e.g., discussion in][]{PZ99,Slesnick08}.
The observed \textit{apparent} luminosity spreads in the data are easily
mis-interpreted as age spreads \citep[see discussion in][]{Hill08}.
It is therefore important to keep in mind that any
{\em observed luminosity spread} in a CMD (or HRD) is
always only an {\em upper limit} to a {\em possible age spread}.
Underestimating the uncertainties in the observational 
data can easily lead to biased
results\footnote{To mention just two
examples we refer to the study of the young cluster NGC~3293
by \citet{BF03}, in which a careful analysis led to a correction of
previous claims about the stellar ages, and to
the new results of \citet{Currie10} for 
h and $\chi$ Persei, that clearly refute earlier claims
of different ages and large age spreads for the double cluster.}.
We will therefore not attempt to determine ages and masses for individual stars,
but restrict our interpretation of the CMDs to estimates of the
{\em typical ages} for {\em total population of young stars} in each cluster;
in a large enough sample, most of the above-mentioned uncertainties are
expected to cancel out statistically.
With these caveats in mind we now take a look at the 
CMDs for individual parts  of the CNC.

\begin{figure*}[h]
\includegraphics[width=18.5cm]{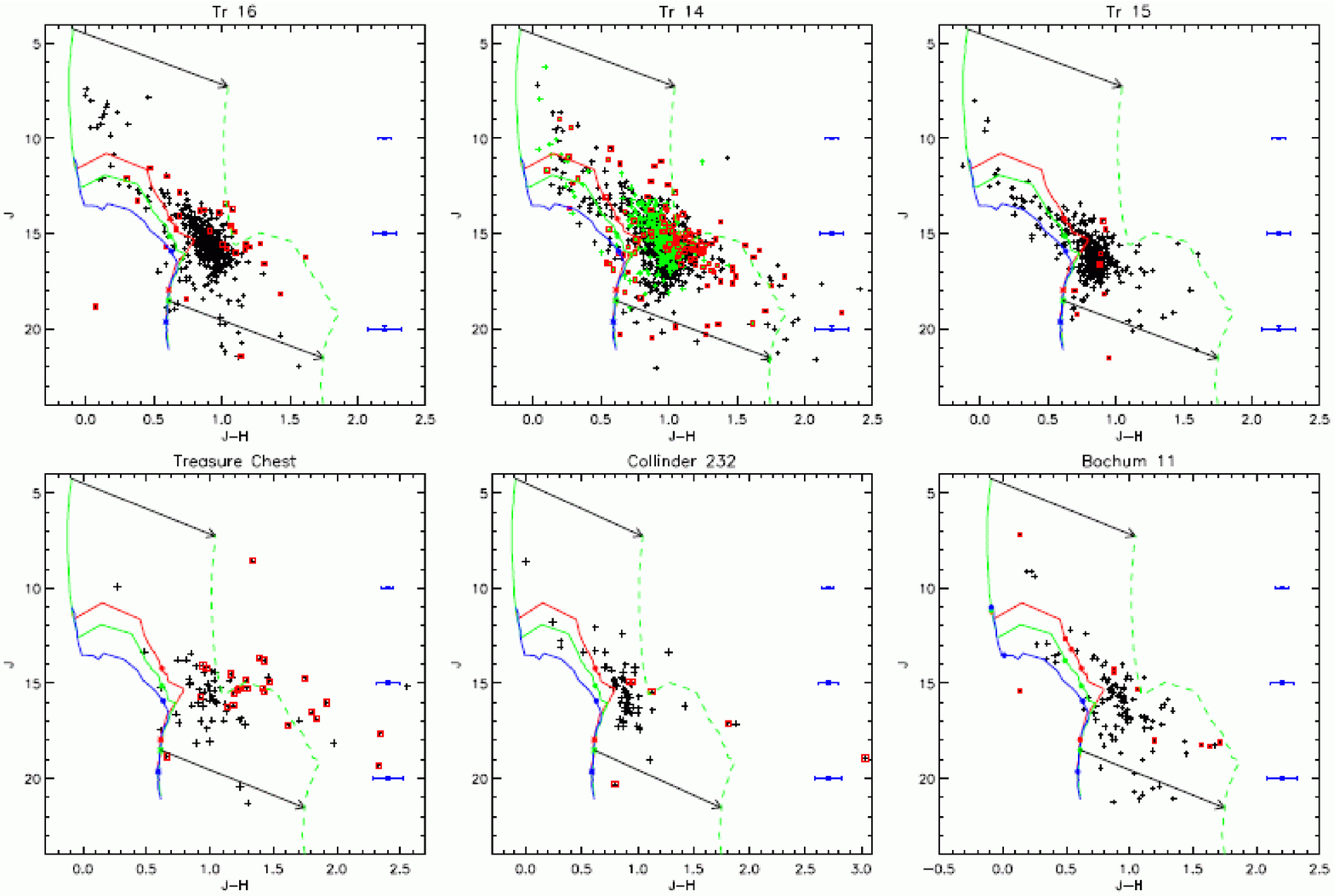}
\caption{Color-magnitude diagrams for the X-ray selected 
Carina members (crosses) in the different clusters in the CNC.
Objects with infrared excesses are additionally
marked by red open boxes.
The solid green line shows the 3~Myr isochrone, the red and the
blue lines show the low-mass ($M \leq 7\,M_\odot$) isochrones
for ages of 1~Myr and 10~Myr, respectively.
The large solid dots mark the positions of $1\,M_\odot$ stars,
the asterisks those of $0.075\,M_\odot$ objects on these
isochrones.
The arrows show the
extinction vector for $A_V = 10$~mag, and the dashed green line
shows the 3~Myr isochrone redenned by that amount.
The sequence of blue errorbars near the right edge
indicates the typical range of magnitude dependent photometric
uncertainties.
In the CMD for Tr~14, the green crosses mark the X-ray sources in the central
$R=2'$ cluster core.
\label{cmds.fig}}
\end{figure*}


\subsection{The cluster Trumpler~16}

The loose cluster Tr~16 is located in the
center of the Carina Nebula and includes the optically dominant
star $\eta$~Car as well as the majority of the O-type stars
in the CNC.
All age estimates for Tr~16 reported so far in the literature concern
only the high- and intermediate mass stars.
\citet{Massey01} found that the
high-mass population of  Tr~16 seems to be co-eval
and report an age of 1.4~Myr (note that they assumed a distance of 3.1~kpc rather
than the value of 2.3~kpc we use here). 
\citet{DeGioia01} claimed that the intermediate-mass stars 
in Tr~16 have been
forming continuously over the last 10~Myr, whereas the high-mass stars
formed within the last 3~Myr.
\cite{Dias02} list an age of 5~Myr, whereas
 \citet{Tapia03} suggested continuous star formation since 6~Myr until now.
These claims of large age spreads may be overestimates caused by
the considerable differential reddening across the areas of these
loose clusters. The most reliable 
age estimate seems to be the value of $\sim 3-4$~Myr that
was suggested by \citet{SB08} based on 
detailed stellar evolution models for $\eta$~Car,
the presence of O3 main sequence stars, and the modeling of the
main-sequence turn-off in optical color-magnitude diagrams.

The CMD of Tr~16 shown in Fig.~\ref{cmds.fig} is based on the
449 stars with complete $J$- and $H$- and $K_s$-band photometry
among the 529 X-ray sources that are classified as Carina members
and associated to one of the sub-clusters in the Tr~16 area
by \citet{Feigelson11}.
Concerning the age of the low-mass stellar population, we note
that our CMD of the X-ray selected sources
shows a conspicuous group of stars with $J \sim 12 - 14$ apparently aligned
with the $3\!-\!4$~Myr isochrones. The bulk of the low-mass stars
is also seen at locations consistent with ages of $\approx 3 - 4$~Myr,
and thus we consider this to be 
the most likely age for the low-mass stellar population 
in Tr~16.
This value is 
consistent with the most likely age of the high-mass members,
suggesting that high- and low-mass stars have
formed at the same time.
We note that a
more comprehensive study of the X-ray selected population 
of Tr~16 can be found in \citet{Tr16-CCCP}.

\subsection{The cluster Trumpler~14}

Tr~14 is the second most massive cluster in the Carina Nebula
and contains 10 known O-type stars.
Its spatial configuration is considerably more
compact than Tr~16. 
The published age estimates for Tr~14 show remarkable
discrepancies: While
\citet{DeGioia01} claim continuous star formation over the last  
10~Myr, \citet{Dias02} list an age of 2~Myr.
\citet{Tapia03} suggested a scenario of continuous star formation
lasting since 5~Myr until less than 1~Myr ago.
\citet{Ascenso07} claimed a more or less continuous
age distribution between zero and $\sim 5$~Myr,
and suggested three
peaks in the age distribution at values of
$0\!-\!0.3$~Myr, $1.4\!-\!2.5$~Myr, and $3.2\!-\!5$~Myr.
\citet{Sana10} analyzed new adaptive-optics images of the
cluster core (radius = $2'$) and derived a very young cluster age 
of only $\sim 0.3\!-\!0.5$~Myr.

Our CMD of Tr~14 (Fig.~\ref{cmds.fig})
is based on the membership to the rather extended X-ray cluster 
as revealed by  \citet{Feigelson11}.
It contains 1378 {\it Chandra} sources, for
1219 of which we have complete NIR photometry.
The CMD suggests
a somewhat broader age distribution than for Tr~16.
Some objects lie
between the 3~Myr and 10~Myr isochrones, but we note that
most of these apparently
older objects are located in the outer parts of the cluster.
If we restrict the sample to the $R=2'$ cluster core, 
most of the apparently
older objects are removed. The tendency for younger
stellar ages in the cluster center
agrees with the recent results from
\citet{Sana10}.
The bulk of the stars in the cluster center seem to be quite young,
$\lesssim 3$~Myr. 
Since this value is in reasonable agreement with the
age estimate for the high-mass stars in Tr~14 by \citet{Dias02},
we conclude that low- and high-mass stars in Tr~14 seem to
be coeval.
Many stars in the outer parts of the cluster seem to be somewhat
older; however, it remains unclear whether these
stars in the outer regions are actually members of the cluster
Tr~14,
or whether they may be part of the widely distributed population
of (partly older) stars in the CNC.

\subsection{The cluster Trumpler~15}

The cluster Tr~15 is located in the northern part of the CNC and
contains 6 O-type stars.
It is thought to be several Myr older than Tr~16 and Tr~14. 
 \citet{Dias02} list an age of 8~Myr, while
\citet{Tapia03} claim 
a range of ages between $\sim 4$~Myr and $\sim 30$~Myr, with
a median value of about 8~Myr for the high-mass stars in Tr~15.

Our CMD (Fig.~\ref{cmds.fig})
is again based on the membership analysis of
\citet{Feigelson11}.
It contains 481 {\it Chandra} sources, 
for 436 of which we have NIR photometry.
The upper part of the CMD is dominated by a
group of stars lying between the 3~Myr and the 10~Myr isochrones;
together with the
clear lack of bright objects to the right of the 3~Myr isochrone,
this suggests an age of $\sim 5-8$~Myr for the 
low-mass stars. 
This agrees with the age derived for the
high-mass stars in Tr~15, suggesting that high- and low-mass stars
are co-eval, and clearly showing
that Tr~15 is older than Tr~16 and Tr~14.
Finally, we note that a
comprehensive study of the X-ray selected population 
of Tr~15 can be found in \citet{Tr15-CCCP}.

\subsection{The ``Treasure Chest'' cluster}

The most prominent of the embedded clusters in the
Southern Pillars region is the 
``Treasure Chest'', which is thought to be
very young ($< 1$~Myr; see Smith et al.~2005).
The most massive member is a O9.5V star.
The X-ray clustering analysis of \citet{Feigelson11} associated 96 {\it Chandra} sources
to the Treasure Chest. The 78 sources for which we have NIR photometry
are plotted in Fig.~\ref{cmds.fig}.
The typical extinction of the member stars,
 $A_V \ge 5$~mag, is
considerably larger than in the other regions.
The CMD of the X-ray selected stars is consistent with a very young age,
since nearly all stars lie well above the 1~Myr isochrone.


\subsection{The cluster Collinder 232}

The clustering Cr~232 is located about $3'$ to the east of Tr~14 and 
defined by the early type stars HD~93250 (O3.5V), 
HD~303311 (O5), and HD~93268 (A2). Until recently, the reality
of a stellar cluster was not clear.
\citet{Tapia03} noted that Cr~232 may be just a random
coincidence of these three relatively bright stars.
However, the spatial distribution of the X-ray detected young stars
in the CCCP
clearly shows a significant density peak in the Cr~232 region
\citep{Feigelson11}. 
There are 70 {\it Chandra} sources associated with this peak, for 
67 of which we have $J$+$H$+$K_s$ photometry.
The CMD is consistent with the assumption that
the majority of the stellar population in this region is very young,
$\lesssim 3$~Myr.

Our \mbox{HAWK-I} images reveal several very red, deeply embedded infrared sources
in Cr~232, including the prominent circumstellar disk object discussed
in more detail in \cite{Preibisch11c}. This detection of
embedded infrared sources, and the jets discussed in Sect.~\ref{jets-sect}
clearly shows that Cr~232 actually
is a cluster of very young stars and with ongoing star formation activity.
We note that the IRAS and MSX data also led to the detection
of deeply embedded luminous
infrared sources in Cr~232 \citep[e.g.,][]{Mottram07}, some of which may be
relatively massive YSOs.

\subsection{The cluster Bochum 11}

Bochum 11  (see Fig.~\ref{bo11.fig}) is a loose open cluster in the 
southeastern part of the CNC (i.e.~in the South Pillars region)
and contains 5 O-type stars.
From optical color-magnitude diagrams, age estimates of
$\le 3$~Myr \citep{Fitzgerald87}  
and $\sim 6$~Myr \citep{Dias02} for the high-mass stars have been published.

The X-ray clustering analysis associated 136 {\it Chandra} sources
to Bochum~11. 
The CMD of the 117 X-ray selected stars with  available NIR photometry
(Fig.~\ref{cmds.fig}) shows a group of stars lying between the 3~Myr and 
the 10~Myr isochrones. This suggests that Bo~11 is somewhat older
than Tr~14 and Tr~16, probably around 5~Myr. This agrees reasonably well
 with the
age estimate for the high-mass stars from \citet{Dias02}.


\subsection{The ``widely distributed'' population}

\begin{figure}[h]
\includegraphics[width=8.5cm]{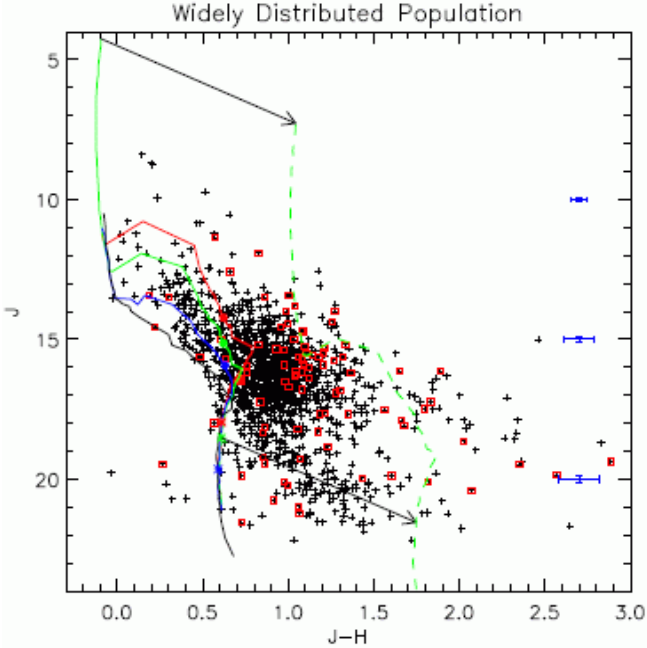}
\caption{Color-magnitude diagram for the X-ray selected 
Carina members in the ``widely distributed population'' within the
HAWK-I survey area.
The meaning of the symbols is as in Fig.~\ref{cmds.fig}.
\label{cmd-wdp.fig}}
\end{figure}

According to the clustering analysis of \citet{Feigelson11},
5185 of the 10\,714 X-ray detected Carina members
 are not associated with any cluster
and thus constitute  a widely dispersed
population of young stars.
Note that due to the spatial sensitivity variations
(caused by the off-axis mirror vignetting and
degradation of the point spread function), the X-ray detection completeness 
for this dispersed population is somewhat lower than for the 
previously known clusters (that were preferentially observed at
small on-axis angles in the CCCP survey mosaic).
This implies that the relative size of this dispersed population
(compared to the X-ray detected populations of the clusters)
is somewhat underestimated.

For 1412 of these objects in the HAWK-I survey area
we have complete $J$- and $H$- and $K_s$-band photometry.
The CMD of these distributed stars (Fig.~\ref{cmd-wdp.fig}) 
is clearly different from the CMDs of the clusters Tr~14, 15, and 16,
because it shows a much broader spread of colors.
The considerable number of objects to the left of the 3~Myr isochrone
and several objects near or to the left of the 10~Myr isochrone
suggest a rather broad age distribution.
We also note that the relative number of bright objects is considerably
smaller than seen in the CMDs of the clusters; this suggests 
that the fraction of high-mass stars is smaller in the
distributed population compared to the clusters.

\section{NIR excess fractions and ages of the different clusters in the CNC}

\begin{figure}
\includegraphics[width=8.5cm]{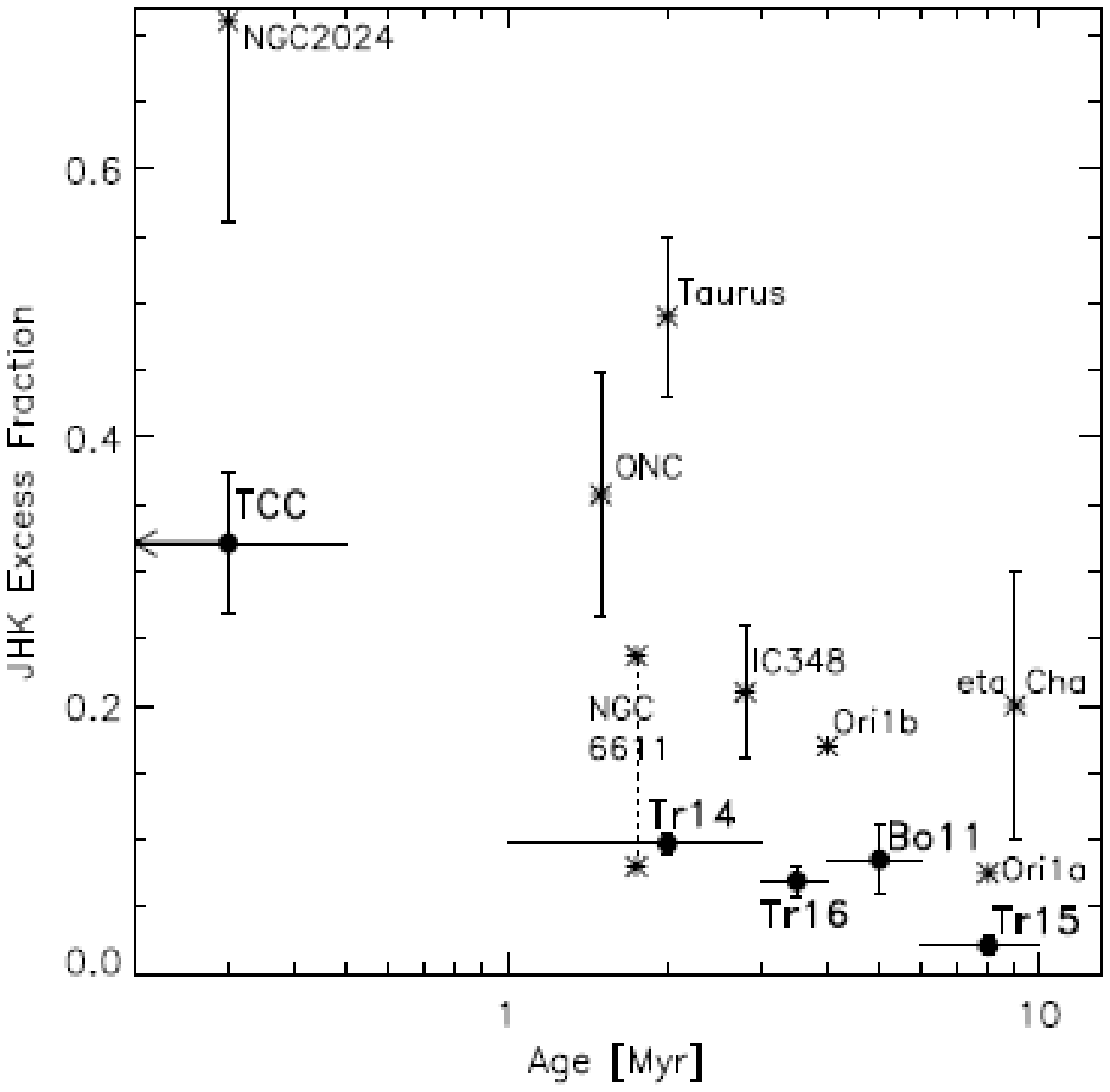}
\caption{JHK excess fraction versus age for the clusters
in the CNC. For comparison, we also show JHK excess fractions
derived for 
NGC~2024 and Taurus \citep{Haisch00},
the Orion Nebula Cluster 
\citep[ONC;][]{Wolk05},
IC~348 \citep{Haisch01}, 
eta Cha \citep{Lyo03},
Ori~OB~1a and 1b \citep{Briceno05}.
For the case of NGC~6611, we plot the spatial variation
 of excess fractions from the
center of the cluster to the outer parts 
as derived by \citet{Guarcello07}.
              \label{excess-age.fig}%
    }
\end{figure}

The fraction of X-ray selected stars with $JHK$ excess emission
has been determined in \citet{Preibisch_CCCP} to be
$(9.7\!\pm\!0.8)\,\%$ in Tr~14, $(2.1\!\pm\!0.7)\,\%$ in Tr~15,
$(6.9\!\pm\!1.2)\,\%$ in Tr~16, and $(32.1\!\pm\!5.3)\,\%$ in the Treasure Chest.
For Bochum~11 we find that 10 of the 117 sources with NIR photometry
have NIR excesses, giving an excess fraction
of $(8.5\!\pm\!2.6)\,\%$. Here we consider the relation between these
excess fractions and the ages of the individual clusters derived
from the analysis of the CMDs as described above.
Figure~\ref{excess-age.fig} shows our results for the clusters in the CNC
and compares it to similarly determined JHK excess fractions
in other galactic clusters.

It is obvious that the NIR excess
fractions for all the clusters in Carina are systematically
and considerably lower (by a factor of about two) than 
those for the other young galactic clusters of similar ages.
We interprete this difference as a consequence of the very harsh conditions
due to massive star feedback in the CNC.
The other clusters used for comparison contain much smaller numbers of 
massive stars and are thus characterized by much more quiescent conditions.
The large population of very massive stars in the CNC
leads to a level of hydrogen ionizing flux that is about 
150 times higher than in the Orion Nebula \citep{SB08}; the resulting  
increased heating and photoevaporation of the
 circumstellar disks around low-mass stars in the CNC is expected to
lead to a quicker dispersal of these disks \citep[e.g.,][]{Clarke07}.

This interpretation is supported by results about the
spatial variation of the NIR excess fraction in the
young cluster NGC~6611, that also contains a considerable population
of massive stars in the center. \citet{Guarcello07}
found that the excess fraction of the X-ray detected
low-mass stars is correlated with the distance
from the massive stars and drops from 24\%  for the outer parts
of the cluster to 8\% 
in the center (i.e.~close to the massive stars). This variation
agrees well with the difference in the excess fraction between
the CNC clusters and the other galactic clusters of similar ages.

\section{The size of the stellar population in the CNC}

Our infrared data of the X-ray selected CNC members provide information
about the masses of these stars that can be used to infer properties
of the stellar mass function in the CNC.
For this, we first have to consider the X-ray detection completeness.
The study of the CCCP data by \cite{Broos11a} shows that
the completeness limit for lightly obscured stars observed at small-to-moderate
off-axis angles on the detector is $L_{\rm X} \approx 10^{30}$~erg/s
in the $0.5-8$ keV band.
Assuming that the young stars in the CNC follow the
relation between stellar mass and X-ray luminosity
established by the data from the 
\textit{Chandra} Orion Ultradeep Project \citep[see][]{Preibisch_coup_orig},
we can expect a detection completeness of $\ge 80\%$ for stars with 
$M_\ast \ge 1\,M_\odot$.
This completeness
drops towards lower masses and should become $\la 50\%$ for 
$M_\ast \leq 0.5\,M_\odot$.

We can thus assume the sample of X-ray selected Carina members to be
nearly complete for stellar masses $M_\ast \ge 1\,M_\odot$.
In order to determine the observed number of stars above this
mass limit, we used the CMD for the X-ray selected
Carina members in the HAWK-I survey area.
We find 3182 objects at CMD positions 
corresponding to $M_\ast \ge 1\,M_\odot$ for an assumed age of 3~Myr
(i.e.~above the reddening vector
originating from the predicted position of 3~Myr old $1\,M_\odot$ stars).

It is interesting to compare this number 
to an IMF extrapolation
based on the known population of high-mass stars
in the CNC.
The compilation
of \citet{Smith06} lists 127 individually identified
members with known spectral types, and 52 of these should have
stellar masses $M \geq 20\,M_\odot$ according to 
the calibrations of \citet{Martins05}.
The recent study of \citet{Povich11b} lead to the detection
of further candidates for young high-mass stars in the CNC and
suggests that the true number of high-mass stars
should  be about $\sim 50\%$ larger than the known population. 
Assuming thus a number
of $\approx 78$ stars with $M \geq 20\,M_\odot$, we can use the
model representation of the field star IMF by \citet{Kroupa02}
and find that the predicted total population of stars with $M_\ast \ge 1\,M_\odot$
is $\approx 4400$. 
Since 80\% of the known high-mass stars are inside the HAWK-I survey
area, we thus have to expect that $\approx 3500$ stars with 
$M_\ast \ge 1\,M_\odot$ should be found
within the HAWK-I survey area, if the IMF in the CNC follows the
Kroupa field star IMF.

This prediction agrees very well to the number of
3182 X-ray detected stars with
 CMD locations corresponding to $M_\ast \ge 1\,M_\odot$.
This agreement 
 directly shows that the number of X-ray detected 
low-mass stars in the CNC is as high
as expected from field IMF extrapolations of the high-mass stellar population.
It is a direct confirmation that the IMF in the CNC is consistent
with the field star IMF down to (at least) $M_\ast \approx 1\,M_\odot$.

This conclusion is further supported by the finding that
the shape of the K-band luminosity function of the X-ray selected Carina members
agrees well with that derived for the Orion Nebula Cluster
\citep{Preibisch_CCCP}. This implies
that, down to the X-ray detection limit around $0.5-1\,M_\odot$,
the shape of the IMF in Carina is consistent with that in Orion 
(and thus the field IMF).

These results 
directly show that there is clearly
{\em no deficit of low-mass stars in the CNC} down to $\sim 1\,M_\odot$. 
This is important because the issue whether the IMF is universal or 
whether there are systematic IMF variations in different environments
is still one of the most fundamental open questions of star-formation theory
\citep[see][and references therein]{Bastian10}.
It was often claimed that some
(very) massive star forming regions have a
{\em truncated IMF}, i.e.~contain much smaller
numbers of low-mass stars than expected from the field IMF.
However, most of the more recent and sensitive studies of massive star
forming regions \citep[see, e.g.,][]{Liu09,Espinoza09}
found the numbers of low-mass stars in agreement with the expectation
from the ``normal'' field star IMF. Our result for the CNC
confirms this and supports the assumption of a universal IMF
(at least in our Galaxy).
In consequence, this result also 
supports the notion that OB associations and
very massive star clusters are the dominant formation sites
for the galactic field star population, as already suggested by
\citet{MS78}.

Based on this  result, we can now proceed and make an estimate of the total
stellar population (i.e.~down to $\approx 0.1\,M_\odot$)  of the CNC
by assuming that the X-ray luminosity function
in the CNC is similar to that in Orion (as suggested by the data).
The study of the CCCP source statistics by
\citet{Feigelson11} found that the 3220 bright X-ray sources with
a photon flux of $\log(F) \ge -5.9$~photons/sec/cm$^2$ in the 0.5--8~keV band
constitute a spatially complete sample. If we 
transform this flux limit according to the difference in distances
(i.e.~415~pc for the ONC versus 2.3~kpc for the CNC), the comparison
to the distribution of photon fluxes in the \textit{Chandra} Orion Ultradeep Project
sample \citep{Getman05}
yields an estimate of $\sim 43\,727$ stars in total for the CNC.
This number
agrees very well with the extrapolation of the
field IMF based on the number of massive stars (see above): 
for 78 stars with  $M_\ast \ge 20\,M_\odot$,
the Kroupa IMF predicts that there should be $\approx 40\,000$ stars
with  $M_\ast \ge 0.1\,M_\odot$.

Finally, we can multiply the extrapolated number of $\sim 43\,727$ stars
with the mean stellar mass of
$0.64\,M_\odot$ (as valid for the Kroupa IMF in the $0.1\,M_\odot$ to $100\,M_\odot$
range),
and find a total stellar mass
of the inferred CNC population
of about $28\,000\,M_\odot$.

\section{Search for faint star clusters and results on the spatial distribution 
of the young stars\label{clusters.sect}}

The spatial distribution of the stars in the CNC contains
important information about the structure, dynamics, and
evolution of the region. 
A very important question in this context is whether the stars form 
in a clustered mode or in a dispersed mode.
Until recently, only the high-mass part
of the stellar population was well known. Most of the massive stars
are in one of the open clusters (and this was the reason why the CNC was often
considered to be a ``cluster of clusters''). 
However, according
to the field IMF, most of the total stellar mass is in the low-mass
stellar population\footnote{According
to the Kroupa IMF, 89\% of the total stellar mass in the $0.1-100\,M_\odot$ range
is contained in stars with $M_\ast < 20\,M_\odot$, 
and 73\% 
in stars with $M_\ast < 5\,M_\odot$.}.
Recent studies have shown that 
a significant fraction of all stars in the solar neighborhood seems to form
in a non-clustered, dispersed mode \citep[e.g.,][]{Gutermuth09}. 
\citet{Bressert10} suggested that the stellar
surface densities of young stars in nearby star forming regions 
follow one single smooth distribution rather than 
two different discrete modes (i.e.~clustered versus distributed).

\begin{figure*}
\includegraphics[width=12cm]{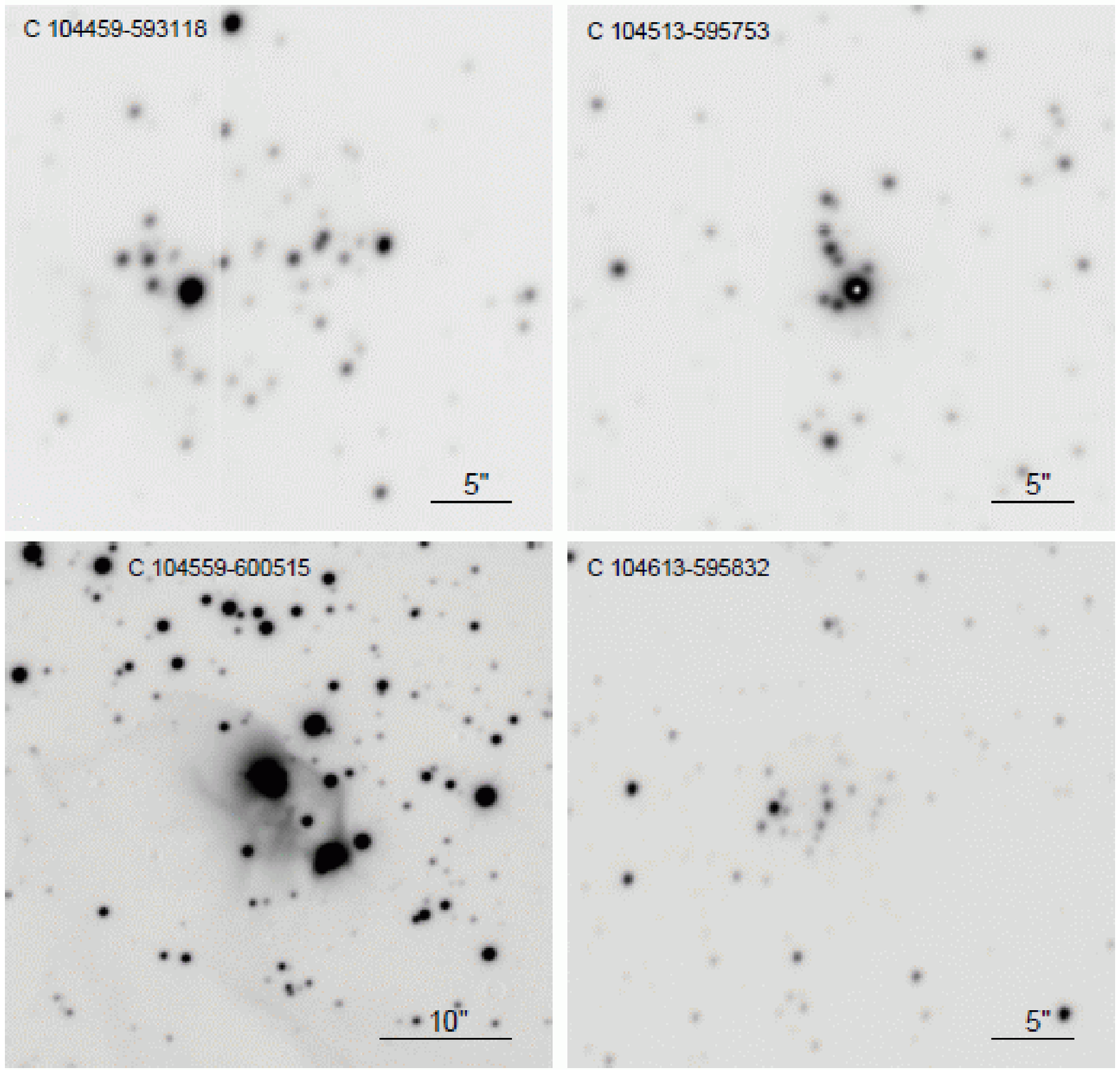}
\caption{HAWK-I $H$-band images of the four
newly detected clusters.
              \label{clusters.fig}%
    }
\end{figure*}

The spatial distribution and clustering properties of the
X-ray detected YSOs in the CNC was studied in detail by
\citet{Feigelson11}.
They identified 20 principal clusters of X-ray stars (most of
which correspond to known optical clusters in the CNC) and
31 small groups of X-ray stars outside the major
clusters.
Altogether, these clusters contain about half of the
X-ray detected YSOs in the CNC. The other half of the
X-ray detected YSO population seems to constitute
a widely dispersed, but highly populous,
distribution of more than 5000 X-ray stars.
A similar result was found in the analysis of the
\textit{Spitzer} data:
most of the $\sim 900$ identified YSO candidates 
are spread throughout the South Pillar region
\citep{Smith10b}.
Although this \textit{Spitzer} study led to the detection of eleven 
previously unknown clusters, the populations of these clusters
are so small (at most $\le 35$ YSOs per cluster) that
only a small fraction of all YSOs are in one of these new clusters.
While the analysis of the \textit{Chandra} and \textit{Spitzer} data
yielded consistent results, the limited sensitivity of both data sets 
may leave some faint clusters undetected.
The X-ray data are known to be seriously incomplete for masses
$\la 0.5\,M_\odot$. The same is true for the  \textit{Spitzer} data,
that have a completeness limit
of $[3.6] \sim 13$, corresponding approximately to
$M_\ast = 1\,M_\odot$.
This implies that  clusterings consisting of only a few dozen
low-mass stars could have easily been missed.
Such clusterings
should, however, be easily visible in the HAWK-I images,
if their spatial configuration is compact enough.
We have therefore performed a detailed visual inspection
of the HAWK-I images to search for yet unrecognized clusters.

\begin{table}
\caption{Parameters of the newly detected clusters. We list the 
apparent radius $R$, the number of stars $N_\ast$,
the number of NIR excess objects $N_{\ast,\,{\rm exc.}}$,
the number of X-ray detected stars $N_{\ast, {\rm X}}$ in each cluster,
and the $K_s$-band magnitude of the brightest star.}
\label{clusters.tab}       
\begin{tabular}{c|ccccc}
Name & $R$ &  $N_\ast$  &   $N_{\ast,\,{\rm exc.}}$ & $N_{\ast, {\rm X}}$ &Brightest star\\
$[J2000]$& [$''$]  &   & & & $K_s\,[{\rm mag}]$\\\hline
C\,104459\,--593118 & 11 & $\approx 33$  &  8  & 2&  10.90\\
C\,104513\,--595753& 6 & $\approx 10$ & 1  & 4 & \,\,9.30\\
C\,104559\,--600515& 8 & $\approx 16$  &3  & 2 & 10.60\\
C\,104613\,--595832& 7  & $\approx 14$  &8  &2  &12.86\\
\end{tabular}
\end{table}

The very inhomogeneous cloud extinction causes
strong spatial variations in the surface density of observed background 
objects in the HAWK-I images;
this limits the ability to recognize loose clusterings consisting
of only a small number ($\la 10$) of stars.
We therefore restricted the search to clusters that consist
(apparently) of at least 10 stars and have a configuration dense enough
for them to clearly stand out from their surroundings.
This search revealed only four likely clusters that were not
known before. Images of these clusters are shown in
Fig.~\ref{clusters.fig} and their properties are listed
in Table~\ref{clusters.tab}.
We note that the physical nature of these apparent clusters is not 
entirely clear; some of them might be just random superpositions of
unrelated objects at different distances. The cluster
C\,104559-600515 is the most reliable, because this group of stars 
appears to be embedded in the
head of a prominent pillar in the South Pillars region.
However, even in this case it remains unknown whether the apparent
cluster is a gravitationally bound physical entity.

Despite the uncertainties about the physical reality of these clusters,
our search provides a clear result: it directly confirms that there
is no significant number of previously undetected clusters.
The HAWK-I images clearly show that most of those young stars in the
complex that are not associated to one of the already known clusters
are in a non-clustered, dispersed spatial configuration.
This result strongly supports the conclusions drawn from the X-ray clustering
study of \citet{Feigelson11} that about half of the total
young stellar population is in a widely distributed
spatial configuration.
It also supports the picture of 
small-scale triggered star formation by radiative (and wind) feedback
from the massive stars (see discussion in Sect.~\ref{conclusions.sec}).

\section{Search for protostellar jets \label{jets-sect}}

\begin{figure*}
\includegraphics[width=17.0cm]{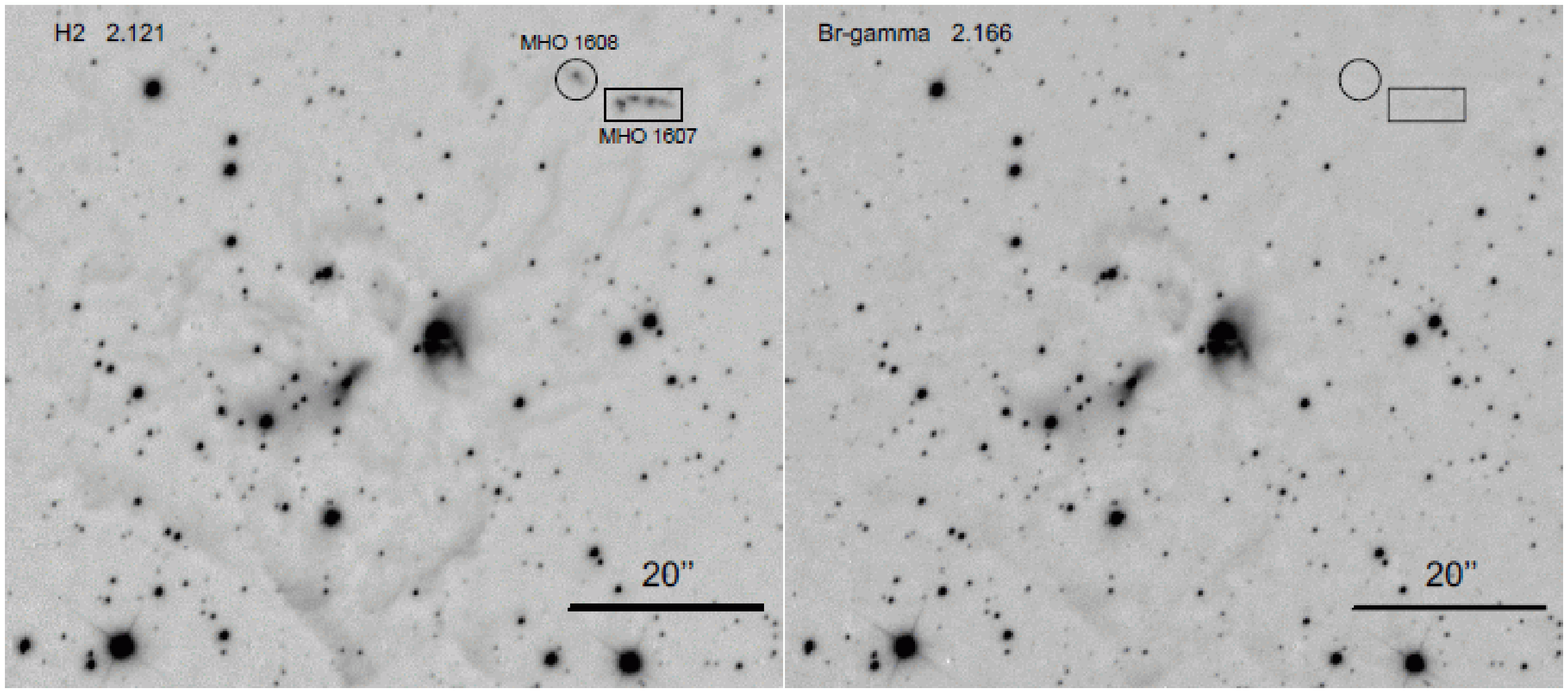}
\caption{Images of the Cr~232 region obtained through the
$2.121\,\mu$m H$_2$ narrow-band filter (left) and the $2.166\,\mu$m Br$\gamma$
narrow-band filter (right). The location of the two molecular hydrogen jets
found in this area, MHO~1607 (104428.2-593245) and MHO~1608 (104429.1-593242),
are marked by the box and the circle in the upper right part of the
images. The bright extended object just right of the center is the
edge-on circumstellar disk object described in detail in 
\citet{Preibisch11c}.
              \label{cr232-jets.fig}%
    }
\end{figure*}

Jets and outflows are important signposts for embedded protostars
that often remain undetected at NIR wavelengths, and
can reveal the latest generation of the
currently forming stars.
When the outflowing material interacts with the environment,
shocks are generated where the flows 
impact with surrounding dense clouds. In these shocks,
hydrogen molecules can be collisionally excited into higher
ro-vibrational levels and the subsequent decay of these excited states causes
strong emission lines in the NIR wavelength range.
The 2.12~$\mu$m  $\nu=1\!-\!0$\, S(1) ro-vibrational emission line of
molecular hydrogen is a very convenient and widely used tracer of these
shocks \citep[e.g.,][]{McCaughrean94,MSmith07,Davis08}.
This was the motivation for obtaining HAWK-I images 
through a narrow-band filter centered on the 2.12~$\mu$m H$_2$ line.

It is, however, important to keep in mind that the detection of 
ro-vibrational line emission is not a proof of the presence of jet-induced
shocks, because molecular hydrogen can also be
excited in higher ro-vibrational levels by the mechanism of UV fluorescence
\citep[e.g.,][]{Black76}.
Since the Carina Nebula  is characterized by very
high levels of ionizing photon fluxes, UV fluorescence will be a very important
excitation mechanism for H$_2$ ro-vibrational emission.
In our HAWK-I data, a distinction between collisional (shock) excitation
and UV fluorescence can be made 
by comparing the strength of the emission seen
in the H$_2$ filter images  to that in the images obtained through another
narrow-band filter centered on the Br$\gamma$ line.
Collisionally (jet-) excited 
molecular hydrogen from jets is not expected to show Br$\gamma$ emission
\citep[see, e.g.,][]{Nisini02,MSmith07,GL10}, whereas 
UV excited (i.e.~irradiated) matter should also show Br$\gamma$ emission.

\subsection{H$_2$ jets in the HAWK-I images \label{mhos.sec}}

We performed a detailed visual inspection of the HAWK-I narrow-band
images to search for features that are bright in the
 H$_2$ filter and much fainter (or invisible) in the Br$\gamma$ filter.
This search revealed only six clear cases of
Molecular Hydrogen Emission-Line Objects (MHOs).
These objects have been included in the 
``Catalogue of Molecular Hydrogen Emission-Line Objects in Outflows from Young Stars''
(http://www.jach.hawaii.edu/UKIRT/MHCat/), where they are listed as
MHO 1605 to 1610.

The two most prominent H$_2$ jets, MHO~1607 and MHO~1608,
are found in the north-western
part of the cluster Cr~232. A comparison of the
H$_2$ image and the Br$\gamma$ image is shown in Fig.~\ref{cr232-jets.fig}.
The fact that these features are bright in the $2.121\,\mu$m  H$_2$ line
but not visible in the Br$\gamma$ line
confirms the nature of collision (i.e.~jet-shock) induced emission.

MHO~1607 (J104428.2-593245) is composed of (at least) five discernable knots
that extend over about $6''$, approximately in east-west
direction. The morphology seems to suggest that the jet is moving
towards the west. The HAWK-I images do not reveal an obvious candidate
for the source of the flow.
The second object,
MHO~1608 (J104429.1-593242), is more compact and less bright.
It seems possible that these two MHOs belong to a single flow system.
In that case, the total linear extent is about $11''$ and corresponds to 
a physical length of (at least) $0.12$~pc.

Fig.~\ref{cr232-jets.fig} shows that the surface of the
clouds surrounding Cr~232 emits in the H$_2$ line and also in the Br$\gamma$ line,
showing that this is fluorescent emission from the ionization
front at the surface of the cloud.
In addition to the presence of the deeply embedded YSOs in this cloud
\citep[see][]{Preibisch11c}, the detection of these jets clearly shows
that Cr~232 is a prominent site of actively ongoing star formation in the CNC.

\begin{figure*} \sidecaption
\parbox{12cm}{
\includegraphics[width=12cm]{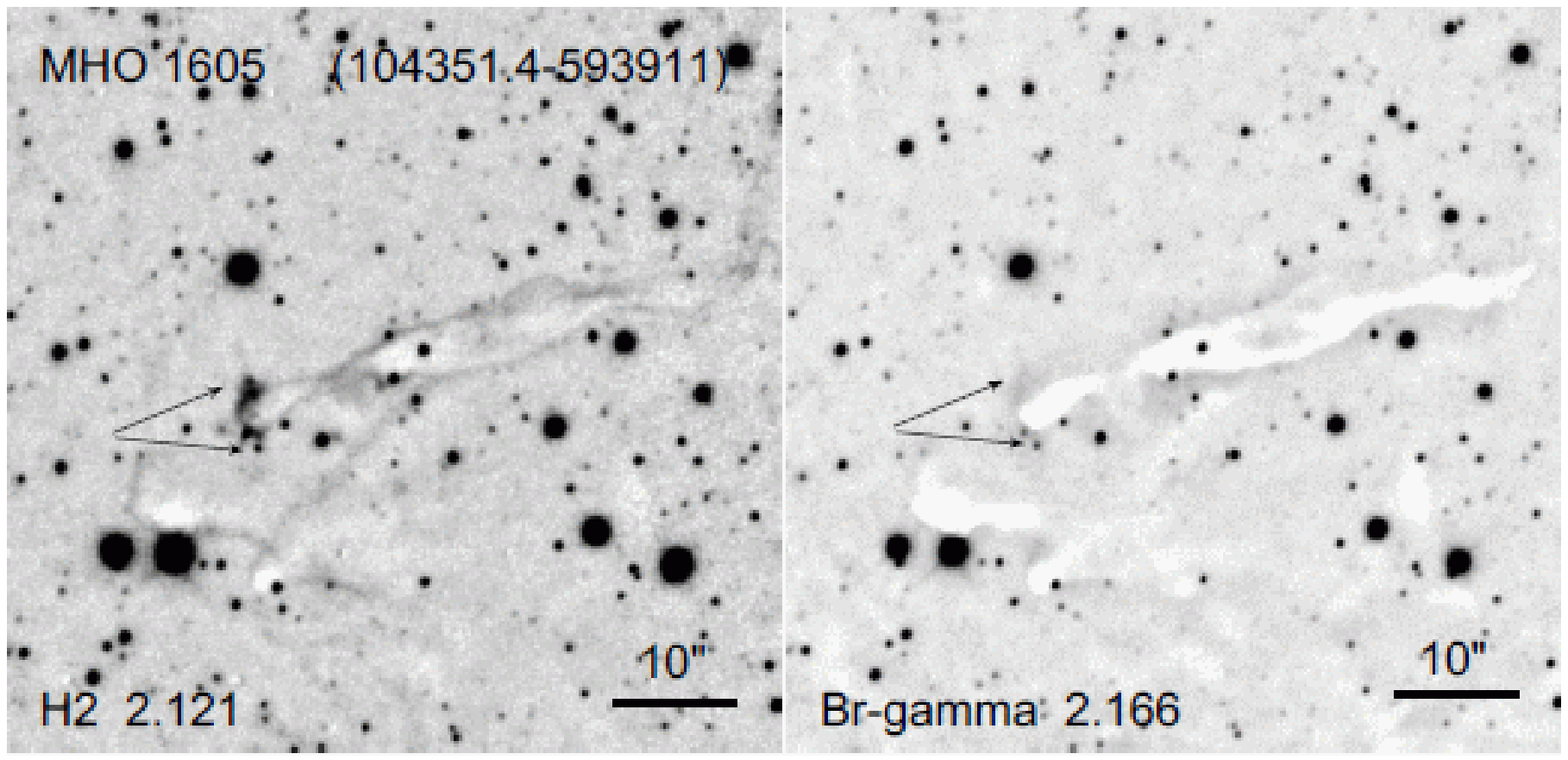}\\
\includegraphics[width=12cm]{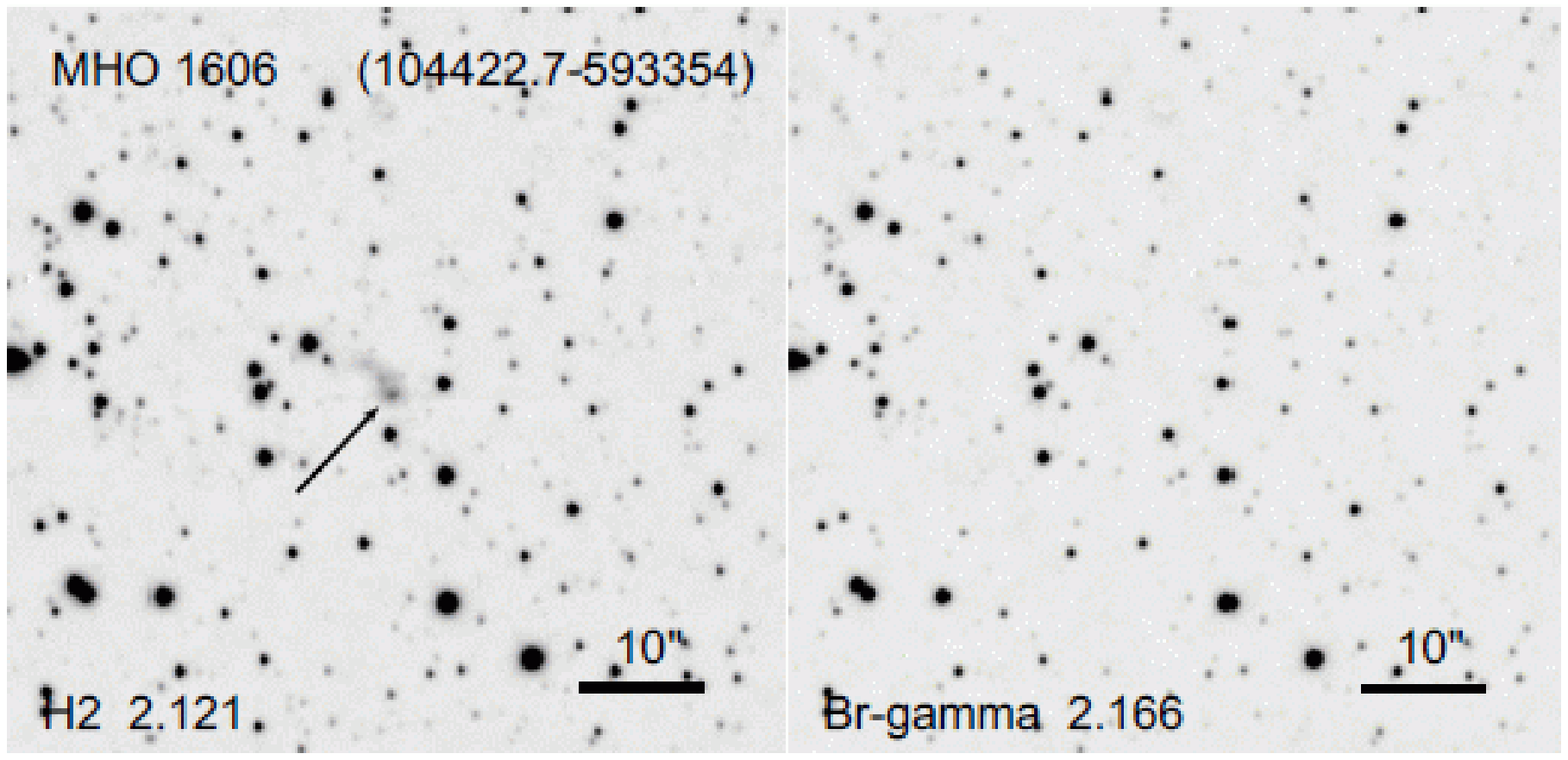}\\
\includegraphics[width=12cm]{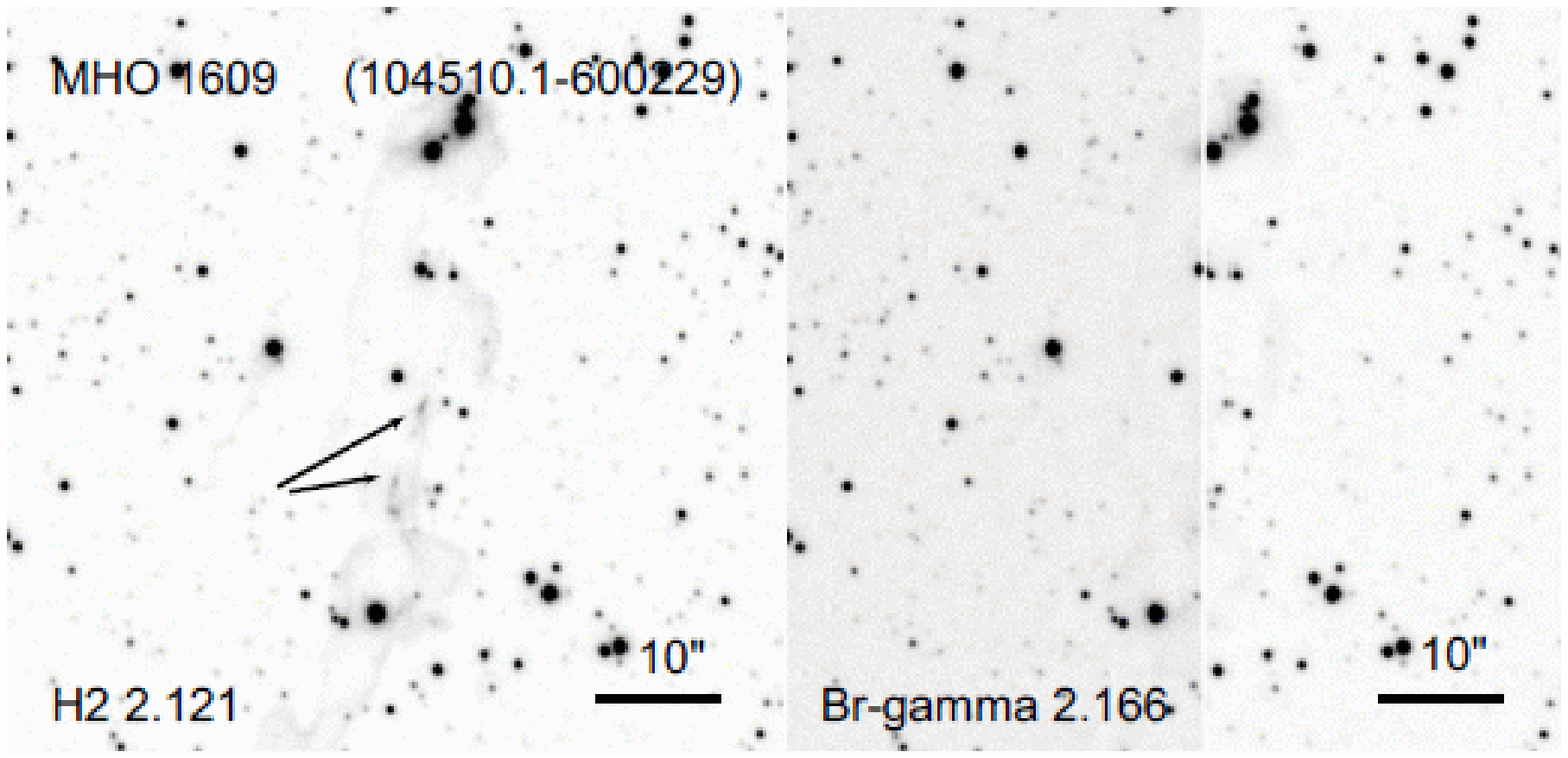}\\
\includegraphics[width=12cm]{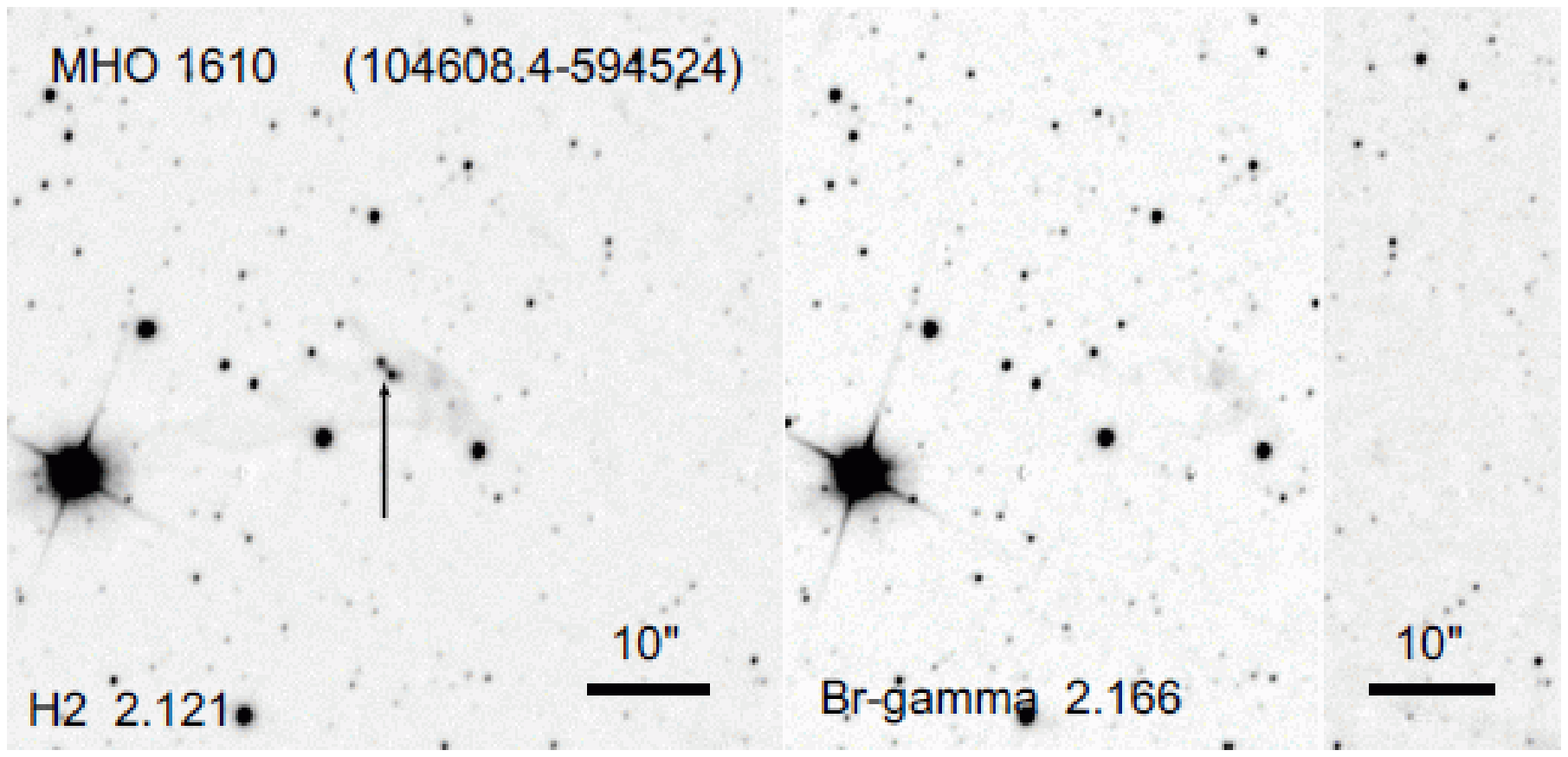}}
\caption{Images of the jets MHO~1005, 1606, 1609, and 1610, obtained through the
$2.121\,\mu$m H$_2$ narrow-band filter (left) and the $2.166\,\mu$m Br$\gamma$
narrow-band filter (right). The location of the molecular hydrogen jets
are marked by the arrows.
              \label{jets.fig}%
    }
\end{figure*}

Figure~\ref{jets.fig} shows the other four MHOs detected in the HAWK-I images.
MHO~1605 (J104351.4-593911) 
is located south of Tr~14. The jet is embedded in the south-eastern tip
of an hour-glass shaped globule. Two patches of H$_2$ emission are seen 
above and below the rim
of the globule, and a faint point-like source is seen inside the globule
and just between them.
MHO~1606 (J104422.7-593354) 
is located south-west of Cr~232. It consists of a 
diffuse knot with a weak $2''$ extension to the north-east.
MHO~1610 (J104608.4-594524) 
is located on the northern rim of a large pillar in the Tr~16 region.
It consists of two
diffuse blobs separated by $1.5''$.

MHO~1609 (J104510.1-600229) 
is located in the middle of a small elongated cloud
in the western parts of the South Pillars region.
This is the only case of a jet found in the central region of
a globule.
The projected length of the flow system is $\approx 10''$, corresponding to
a physical length of (at least) $0.11$~pc. We note that
this cloud is the source of the five optical jet candidates
HHc-4 to HHc-8 described by \citet{Smith10a}, but the MHO seems to be
unrelated to these optical HH objects.
On the other hand, for none of the optical HH object candidates
do we see any molecular hydrogen emission.
The HAWK-I images show a faint diffuse point-like source that may be related to HHc-6,
but since
this object is similarly bright in the H$_2$ and Br$\gamma$ images,
what we see is not H$_2$ line emission but probably reflected light
(perhaps the envelope of the jet-driving protostar).

\subsection{Search for H$_2$ emission from the optical HH objects}

The HST H$\alpha$ imaging survey of \citet{Smith10a} revealed 39 HH jets 
and  jet candidates from YSOs that are embedded in dense globules. 
13 of these HH jets and 14 jet candidates
are located in the area covered by our HAWK-I narrow-band images. Our detailed inspection
of the surroundings of the optical jets in the HAWK-I images showed that
several of the optical HH jets can also be seen in the
NIR  narrow-band images, but in all cases the apparent brightness
in the Br$\gamma$ line filter is similar to that in the H$_2$ line filter.
This implies that the emission we see in the H$_2$ filter image is \textit{not} shock-induced
but either related to UV fluorescence or just reflected continuum light. 
This is most likely a consequence
of the fact that the optical HH jets seen by HST are \textit{irradiated atomic jets}
that have left the globules and expand in the surrounding
diffuse atomic interstellar medium. This implies that no molecular hydrogen should
be present in the immediate surroundings of the HH jets, and thus no H$_2$ line
emission is expected.

However, one might expect to see H$_2$ line emission from those parts of the
jet flow that are still within the dense globules (i.e.~the part of the
path \textit{before} the jets
emerge into the surrounding diffuse atomic medium).
Therefore we performed a detailed inspection of those parts of the globules
from which the optical HH jets emerge.
If the jet-driving protostars were located deeply inside these globules,
one would expect to see H$_2$ line emission within the globule that should be
aligned with the 
optical HH emission outside the surface of the globule. However, 
in no case do we find detectable H$_2$ line emission at such locations.

This null result is in agreement with the fact that none of the optical
HH jets
exhibited detectable excess $4.5\,\mu$m emission (which is also related
to ro-vibrational lines from molecular hydrogen) in the \textit{Spitzer}
imaged analyzed by \citet{Smith10b}.
This has an interesting implication:
it shows that the driving sources of these jets are
located very close to the edge of the globules, and not in the
center of the globules (because then we would expect to see
MHOs). For a protostar located at the
edge (or at the tip) of a globule, the jets will immediately
move into the diffuse atomic medium, and no MHOs are thus expected.
%

\subsection{Implications on the star formation mechanism}

We find only a very small number of six MHOs, i.e.~signposts of protostellar jets, in the
molecular clouds in the CNC. 
The clear lack of H$_2$ jet detections at the positions of the numerous optical HH jets
suggests that the driving sources
are located very close to the surface of the globules, and the
jets therefore move in the atomic gas, not through the molecular gas
in the globules.
This implies that almost all jet-driving protostars (i.e.~currently forming stars)
are located very close to the irradiated edges of the clouds, and \textit{not} in
the central regions of the clouds.

This spatial configuration strongly supports the scenario 
that star formation is currently \textit{triggered} by the advancing ionization
front in the irradiated clouds \citep[see][]{Smith10b}.
This scenario predicts that stars form very close to the irradiated edge
of the clouds, where the radiative compression leads to collapse and
triggers star formation.
In the alternative scenario, if one would assume \textit{spontaneous} star
formation taking place at random locations within these clouds,
most protostars should be located in the inner parts of the
clouds and we would thus expect to see
more cases of MHOs within the clouds, which is not the case.

\section{Summary and Conclusions \label{conclusions.sec}}

\subsection{Summary of the main results}

The analysis of the infrared properties of the 
young stars in the CNC provides important information
about the stellar ages and masses, as well as the mass function
and the total size of the stellar population.
The ages estimated for the low-mass populations
in the young clusters within the CNC are consistent with previous
age determinations for the 
   massive cluster members, suggesting that the high- and low-mass stars 
have the same age, i.e.~have formed together at the same time.

The number of X-ray detected stars with CMD positions
corresponding to stellar masses of $M_\ast \ge 1\,M_\odot$
is  consistent
with an extrapolation of the field star IMF  based on the
number of high-mass ($M \geq 20\,M_\odot$) stars.
This suggests that the IMF in the CNC is consistent with the
field star IMF (down to at least $1\,M_\odot$). 
The extrapolation of the
X-ray detected population down to $0.1\,M_\odot$ suggests a total
population 
of $\approx 43\,730$ stars with an integrated mass of about $28\,000\,M_\odot$. 
We note that this extrapolation only considers the star numbers in 
the X-ray detected, lightly obscured population of young stars in the CNC. 
The analysis of deep \textit{Spitzer} images of the CNC by
\citet{Povich11a} revealed a population of 1439 YSO candidates
with strong mid-infrared excesses that is thought to be 
dominated by intermediate-mass ($2 \dots 10\,M_\odot$) stars.
Most (72\%) of these are {\em not} detected
in the {\it Chandra} X-ray data, probably due to their strong obscuration. 
The extrapolations of \citet{Povich11a} suggests a total population
of $\sim 14\,300$ obscured objects, that have to be added to the
X-ray selected population.
This raises the total population of the CNC to
$\approx 58\,000$ stars, and the integrated stellar mass
 to $\approx 37\,000\,M_\odot$.
These numbers\footnote{As a historical note, 
it is interesting to compare this number to older estimates based on
optical observations only: for example,
\citet{Feinstein95} estimated the total stellar mass of the CNC to be
$3596\,M_\odot$.
The progress in observations has lead to an increase by more than a factor of 10
within just about 15 years!}
clearly show that the CNC constitutes one of the most massive
galactic clusters/associations, probably topping the often quoted
galactic ``starburst templates'' NGC~3603 
\citep[$M_{\ast, \rm tot} \sim 15\,000\,M_\odot$; see][]{Rochau10}
and the Arches cluster 
\citep[$M_{\ast, \rm tot} \sim 20\,000\,M_\odot$; see][]{Espinoza09},
and being similar to
Westerlund~1 \citep[$M_{\ast, \rm tot} \sim 49\,000\,M_\odot$; see][]{Gennaro11}.
The
CNC can  thus be regarded as the most nearby galactic
cluster on the verge of extragalactic starburst clusters.

The presence of several deeply embedded
young stellar objects and the detection of jets 
show that the cloud associated to the cluster Cr~232
is a center of ongoing star formation activity.

The fact that we find only four new small clusters in the HAWK-I images
confirm and extend previous results showing that about half of the
total stellar population in the CNC is in a non-clustered, widely dispersed
spatial mode. The presence of this large distributed stellar population
suggests that the CNC should be considered as an unbound OB association
rather than a cluster of clusters. Since most of the existing clusters are quite
loose and will probably disperse within a few 10~Myrs, the fraction
of widely distributed, non-clustered stars will even increase with time.

The very small number of H$_2$ jets we see in the clouds and the
complete lack of molecular hydrogen emission related to the
numerous known optical HH jets show
that most of the very recently formed protostars
must be located very close to the surface of the globules (not inside), suggesting
a triggered star formation mechanism.

\subsection{Interesting aspects of the current star formation activity 
in the CNC\label{conclusions}}

The UV radiation and winds of the very hot and luminous early O-type 
and WR stars in the central area of the CNC 
affect the surrounding clouds strongly.
While much of the cloud mass was (and still is) dissolved 
and streams away in the expanding super-bubble,
the remaining molecular cloud material is now highly fragmented
into numerous dense pillars. 
The total (dust + gas) mass of the dense clouds in the CNC 
is still quite large ($\sim 60\,000\,M_\odot$), but these clouds
fill only a small fraction of the volume of the CNC \citep[see][]{Preibisch_Laboca}.
 The radiative compression of these clouds
currently leads to  triggered star formation. The spatial distribution of the YSOs
detected in the \textit{Spitzer} survey \citep{Smith10b} suggests that
the induced formation of stars happens predominantly near the most strongly
irradiated tips of the clouds. 
This is consistent with recent results from numerical simulations
of the evolution of irradiated clouds \citep[e.g.,][]{Gritschneder10}.
An implication of this triggered star formation mechanism is that
the cloud material surrounding the newly formed protostars is
simultaneously dissolved by the advancing ionization front.
As their natal surrounding clouds and protostellar envelopes
are dispersed very quickly, these newly formed stars are immediately
exposed to the very harsh radiation field created by the luminous O-type stars,
and therefore loose their circumstellar material very quickly.
The transition from an embedded protostellar object to a revealed
disk-less star thus happens faster than in star forming regions
with lower levels of feedback. This can explain
the very small fraction of young stars with NIR excesses 
as a tracer of circumstellar disks.

Several aspects of this process of triggered
formation of a new stellar generation in the CNC appear particularly 
interesting and are worth being discussed in more detail.
First, the triggering process has produced many sparse groups of 
young stars instead of a few rich clusters; it is thus a 
qualitatively different mode of star formation than
the process that formed the earlier generation of stars in the
much more populous clusters Tr~16 (and Tr~14).
The relatively small sizes of the newly formed clusterings of young stars
suggests that the currently ongoing radiative triggering of star formation
is a very local, small-scale process. This is probably related to the
highly fragmented structure of the now existing dense clouds
\citep[as can be seen in the sub-mm map in][]{Preibisch_Laboca}. The compression
by the ionization front
drives theses rather small individual clouds at various locations in the nebula
into collapse, each of which then gives birth to a small stellar group.

The second aspect is that 
the masses of the stars in the triggered new generation 
seem to be restricted to $\lesssim 20\,M_\odot$, i.e.~to much lower values
than the very massive stars in the older, triggering generation 
($M \ga 100\,M_\odot$ for $\eta$~Car and the O3 and WR stars in Tr~16 and Tr~14).
The infrared and X-ray surveys revealed several candidates for intermediate-mass
and moderately high-mass YSOs, but no trace of any very massive ($M \ge 50\,M_\odot$) protostar 
has yet been found.
This is again supported by the results of the sub-mm study of the structure of the
dense clouds \citep{Preibisch_Laboca}, where it was found that
 nearly all dense clouds in the CNC
have masses $\le 1000\,M_\odot$; according to the empirical relation between
cloud mass and maximum stellar mass, these clouds are expected 
to yield  maximum stellar masses of no more than $\la 15\,M_\odot$.

The third aspect concerns the
size of the triggered new stellar population. The X-ray survey showed that
the widely distributed population consists of a similar number
($\sim 5000$ X-ray detected objects) of stars 
as the populations of the
triggering populations in Tr~16 $+$ Tr~14.
Considering the spatial
incompleteness effect discussed in Sec.~3.7, the dispersed population
is probably even somewhat larger than the cluster populations.
Evidence for a second generation of star formation triggered
by radiative and wind-feedback from the massive stars in a first generation 
has been found in several other massive star forming regions, but
in most cases the total size of the second stellar
generation is considerably smaller than that of the
first generation
\citep[see, e.g.,][]{Stanke02,Reach04,Linsky07,Wang10}.
The remarkably large size of the latest stellar generation in 
the CNC may be related to the particularly high level
of massive star feedback in the CNC, that is perhaps
more effective in triggering star formation
than the much lower feedback levels in regions with smaller
populations of very massive stars in the older generation.

The combination of the deep HAWK-I NIR images with the
results of the X-ray survey
has provided us with a very substantial amount of crucial new information
about the stellar content, the star formation history, and the
ongoing star formation process in the CNC.
Future studies will
combine these observational data with detailed 
numerical simulations of how molecular clouds
evolve under the influence of strong massive star feedback
\citep[see][]{Gritschneder10}.
Such a comparison can provide new and detailed insights
into fundamental
processes such as the disruption of molecular clouds by
massive stars, the origin of the observed complex pillar-like structures
at the interfaces between molecular clouds and \ion{H}{II} regions,
the effect of stellar feedback on molecular cloud dynamics and turbulence,
and how ionizing radiation and stellar winds
trigger the formation of a new generation of stars.

\begin{acknowledgements}
We would like to thank the referee for a very competent review and 
several suggestions that helped to improve this paper.
We thank the ESO staff (especially Markus Kissler-Patig and
 Monika Petr-Gotzens) for performing the HAWK-I observations
in service mode.
We gratefully acknowledge funding of this work by the German
\emph{Deut\-sche For\-schungs\-ge\-mein\-schaft, DFG\/} project
number PR 569/9-1. Additional support came from funds from the Munich
Cluster of Excellence: ``Origin and Structure of the Universe''.
R.R.K.~is supported by a Leverhulme research project grant (F/00 144/BJ).
\end{acknowledgements}

\end{document}